\documentclass[aps, prb, reprint, superscriptaddress, twocolumn, showkeys, amsmath, amssymb]{revtex4-1}
\usepackage[english]{babel}
\usepackage{amsmath, amssymb, bbm, mathrsfs, bm, braket, color, graphicx, comment, amsfonts, dsfont}
\usepackage[colorlinks,citecolor=blue,urlcolor=blue]{hyperref}
\usepackage[mathscr]{euscript}

\newcommand{\tr}{\mathop{\mathrm{tr}}}

\begin{document}
\title{Phase tunable second-order topological superconductor}

\author{S. Franca}
\affiliation{IFW Dresden and W{\"u}rzburg-Dresden Cluster of Excellence ct.qmat, Helmholtzstr. 20, 01069 Dresden, Germany}

\author{D. V. Efremov}
\affiliation{IFW Dresden and W{\"u}rzburg-Dresden Cluster of Excellence ct.qmat, Helmholtzstr. 20, 01069 Dresden, Germany}

\author{I. C. Fulga}
\affiliation{IFW Dresden and W{\"u}rzburg-Dresden Cluster of Excellence ct.qmat, Helmholtzstr. 20, 01069 Dresden, Germany}

\date{\today}

\begin{abstract}

Two-dimensional second-order topological superconductors (SOTSCs) have gapped bulk and edge states, with zero-energy Majorana bound states localized at corners. Motivated by recent advances in Majorana nanowire experiments, we propose to realize a tunable SOTSC as a two-dimensional nanowire array. We show that the coupling between the Majorana modes of adjacent wires can be controlled by phase-biasing the device, allowing to access a variety of topological phases. We characterize the system using scattering theory, which provides access to its transport properties and its topological invariants. The setup is robust against disorder, both in the nanowires themselves and in the Josephson junctions formed between adjacent wires. Further, we identify a parameter regime in which an initially trivial system is rendered topological upon adding disorder, providing an example of a second-order topological Anderson phase.
\end{abstract}

\maketitle

\section{Introduction}

The hallmark characteristic of topological phases of matter is the presence of robust zero-energy modes at the interfaces between topologically trivial and nontrivial regions. The existence of these modes is the consequence of a nonzero, quantized value of the topological invariant, a relationship known as the bulk-boundary correspondence. Until recently, this correspondence only predicted the existence of gapless states with dimension lower by one than that of the bulk. The stability of these gapless modes is ensured by the presence of local symmetries such as the particle-hole, time-reversal or chiral symmetry.\cite{Hasan2010,Qi2011,BernevigBook} In the presence of crystalline symmetries, this correspondence holds only for the surfaces left invariant by these symmetries. \cite{Fu2011,Hsieh2012,Xu2012}

Recently, research focus has shifted towards higher-order topological phases,~\cite{Khalaf2018,Geier2018,Trifunovic2018, Benalcazar2017, Benalcazar2017a, Song2017, Schindler2018, Schindler2018a, Langbehn2017,Wang2018a,Zhu2018,Ezawa2018, Ezawa2018a, Ezawa2018d,Dwivedi2018, Miert2018, Ezawa2018c, Hsu2018, Yan2018, Wu2018,Zhu2018a,Wang2018b,Liu2018,Peng2018,Serra-Garcia2018,Xie2018,Peterson2018,Serra-Garcia2018a,Imhof2018,Zhang2018,Franca2018,Calugaru2019,Wang2018,Benalcazar2018,Attig2018,Queiroz2018,Liu2018b,Araki2018,Ezawa2018f,Ezawa2018e,
Ezawa2018b,Ahn2018,Bultinck2018,You2018,Liu2018a,Bomantara2018,Rodriguez-Vega2018,Huang2018,Volpez2018,Hayashi2018,Noh2018, Agarwala2019,Kheirkhah2019,Slager2015,Ghorashi2019} in which both the D-dimensional bulk and the (D-1)-dimensional boundary are gapped, whereas topological zero modes have a dimension of (D-2) or less. A D-dimensional, N$^{\rm th}$ order topological phase is defined as hosting boundary modes of dimension (D-N), localized, for instance, at the corners or hinges of a sample. To date, higher-order phases are divided into two groups, labeled \textit{intrinsic} and \textit{extrinsic}.\cite{Geier2018, Trifunovic2018} The former ones require lattice symmetries and are termination independent, as long as the termination is compatible with these symmetries. The lattice symmetries are essential, as they impose unique and opposite mass terms on the neighboring surfaces, thus localizing gapless modes at the corresponding hinges or corners. This is quantified by a nontrivial bulk invariant. On the other hand, extrinsic phases do not require any spatial symmetries. In essence, the bulk is trivial and boundary modes exist as a consequence of a nontrivial surface.\cite{Trifunovic2018}

A majority of proposals for the realization of higher-order topological insulators rely on the presence of crystalline symmetries. For instance, Refs.~\onlinecite{Benalcazar2017,Benalcazar2017a} discussed a two-dimensional/three-dimensional (2D/3D) Su-Schrieffer-Heeger model,\cite{Su1979} where a topological invariant, protected by mirror symmetries, was interpreted as a bulk quadrupole/octupole moment. So far, this model has been realized in meta-materials,\cite{Serra-Garcia2018,Xie2018} as well as in microwave and electrical circuits.\cite{Peterson2018,Serra-Garcia2018a,Imhof2018} Furthermore, higher-order topology was used to explain gapless modes along the step edges of a bismuth crystal,\cite{Schindler2018a} protected by the combination of time-reversal symmetry and spatial symmetries like three-fold rotation and inversion. 

Topological superconductors (TSCs) also represent a promising venue to achieve higher-order phases. Some proposals rely on $p$-wave superconductivity,\cite{Langbehn2017,Wang2018a,Khalaf2018,Zhu2018} where a bilayer of $p_x + i p_y/p_x -i p_y$ superconductors has gapped edges due to the application of a time-reversal symmetry breaking term,\cite{Khalaf2018,Zhu2018} or the coupling between the layers is such that certain edges are gapped.\cite{Langbehn2017} Other approaches are based on mixed pairing (usually $ p + i d$ or $s+i d$),\cite{Wang2018a,Wu2018,Zhu2018a} and several realizations have been proposed, including $\rm Sr_2RuO_4$ under certain conditions. Furthermore, SOTSC phases have been proposed in engineered systems like a 2D topological insulator placed in the proximity of a high-temperature superconductor,\cite{Wang2018b,Liu2018} or a combination of a 3D anti-ferromagnetic topological insulator and a conventional s-wave superconductor.\cite{Peng2018} 
An additional advantage of higher-order TSCs is their potential use as generators of quantum codes.\cite{You2018}
Finally, a recent work of Volpez \textit{et al.} discusses a SOTSC phase induced once the in-plane Zeeman field is present in $\pi$-junction Rashba layers.\cite{Volpez2018}

The experimental realization of topological superconductivity has been an active field of research in the last decade.\cite{Qi2011,Beenakker2013} Initially, it was predicted in $p$-wave superconductors, leading to proposals based on the proximity introduced superconductivity in topological insulators\cite{Fu2008,Qi2010} and semiconductors,\cite{Oreg2010,Lutchyn2010} both with strong spin-orbit coupling. Among these, the Majorana nanowire, made of a superconductor deposited or epitaxially grown on a nanowire has been the focus of intense experimental efforts.
\cite{Mourik2012,Deng2016,Zhang2018a, Albrecht2016, Krogstrup2015, Gazibegovic2017, Casparis2018, Vaitiekenas2018a}

Motivated by the recent advances in the fabrication techniques and characterization methods of Majorana nanowires, we use them as a platform to create an extrinsic SOTSC. The 2D array of coupled Majorana nanowires is already studied in the context of weak-TSCs,\cite{Seroussi2014,Fu2007,Fu2007a,Diez2014} where a set of equally spaced nanowires is proximitized by a single superconductor. Here, we propose a simple design by which the array realizes a SOTSC once the inter-wire couplings are dimerized. Such a dimerization could be the result of an alternating wire spacing, which would remain fixed once the device has been fabricated. Alternatively, we show that a continuously tunable setup is realized if the wires are equally spaced, but proximitized by different superconductors. By controlling the superconducting phase differences between adjacent wires, it is possible to tune the coupling between adjacent Majorana modes such that the array forms a SOTSC. We characterize the system using the scattering matrix approach, which is discussed in the context of second-order topological phases by Geier \textit{et al}.\cite{Geier2018} Finally, our numerical simulations show that this phase is robust to disorder, and can, in fact, expand in the presence of it. Therefore, our model realizes an example of a secoSzczecinnd-order ``topological Anderson phase''.\cite{Li2009, Groth2009, Song2012, Lv2013, Adagideli2014}

The rest of the paper is organized as follows. In Section \ref{sec:models}, we do a step-by-step construction of the SOTSC, starting from the model of a single Majorana nanowire. In Section \ref{sec:characterization}, we characterize the system using scattering matrices, determining its topological invariant and transport properties. Section \ref{sec:disorder} is devoted to the effects of disorder on the system. Finally, in section \ref{sec:conclusion}, we summarize our results and draw comparisons to other proposals. 
 
\section{Models} \label{sec:models}

In this Section, we show how the coupling between Majorana bound states (MBS) belonging to adjacent wires may be tuned by varying the superconducting phase difference, and then exploit this tunability to generate a SOTSC phase in a nanowire array.

\subsection{The model of a Majorana nanowire}

We begin by reviewing the model of a single nanowire oriented along the x-direction and proximity coupled to an s-wave superconductor. Only the nanowires with strong Rashba spin-orbit coupling (SOC) are relevant as the SOC, once a magnetic field is applied, creates an effectively spinless state, thus mimicking the physics of the Kitaev chain.\cite{Kitaev2001} 
Then, the Bogoliubov-de Gennes (BdG) Hamiltonian $H_{\rm wire} = \int d k_x \Psi^{\dagger} \mathcal{H}_{\rm wire} \Psi$ is defined through the Hamiltonian density 
\begin{align} \label{eq:MFHam}
\begin{split}
& \mathcal{H}_{\rm wire} (k_x) {} = [2t_x(1- \cos{k_x})  -\mu]\tau_z \sigma_0  + \\ 
&   V_x \tau_0 \sigma_x  +  V_z \tau_0 \sigma_z +  \Delta \tau_x \sigma_0  + \alpha \sin{k_x} \tau_z \sigma_y,
\end{split}
\end{align}
in the basis $\Psi^{\dagger} = (\psi_{k_x,\uparrow}^{\dagger}, \psi_{-k_x \downarrow}^{\dagger}, \psi_{k_x\downarrow}^{}, -\psi_{-k_x\uparrow}^{})$, where $\psi_{k_x, \uparrow}^{\dagger}$ denotes the creation operator of an electron with spin up and momentum $k_x$. As this basis is assumed hereafter, in the following we specify only the Hamiltonian densities.

Furthermore, $t_x$ and $\alpha$ are the nearest-neighbor hopping strength and the SOC in the x-direction (along the wire), $\mu$ is the chemical potential, $V_x/V_z$ is the Zeeman energy in the x-/z-direction, and $\Delta$ is the superconducting pairing strength. Pauli matrices $\tau$ and $\sigma$ represent particle-hole and spin degrees of freedom, respectively.
This system belongs to class BDI of the Altland-Zirnbauer classification,\cite{Altland1997} with particle-hole symmetry $\mathcal{P} = \tau_y \sigma_y \mathcal{K}$, chiral symmetry $\mathcal{C} = \tau_y \sigma_y$ and time-reversal symmetry $\mathcal{T} = \mathcal{K}$, where $\mathcal{K}$ denotes complex conjugation.

From the eigenvalue equation, one obtains
\begin{equation*}
\begin{split}
 & E^2 = \epsilon_{k_x}^2 + V_x^2 + V_z^2+ \Delta^2 + \alpha^2 \sin^2{k_x} \\
 & \pm 2 \sqrt{(V_x^2 + V_z^2) \Delta^2  + (V_x^2 + V_z^2 + \alpha^2 \sin^2{k_x}) \epsilon_{k_x}^2},
\end{split}
\end{equation*}
where $\epsilon_{k_x} = 2t_x(1- \cos{k_x})  -\mu$.

To achieve the topological phase, it is sufficient that only one of the Zeeman terms is nonzero. The external magnetic field applied in the x-direction gives rise to a Zeeman term $V_x$, while $V_z$ is more experimentally relevant for ferromagnetic atoms deposited on an s-wave superconductor,\cite{Nadj-Perge2013,Nadj-Perge2014} where it arises due to the local magnetic moments of atoms. 

The topological criterion, $V_x^2 + V_z^2 > \Delta^2 + \mu^2$, that distinguishes a topologically trivial and a nontrivial phase, is obtained by setting $k_x = 0$. In the case of only one wire, regimes $V_x\ne 0;V_z =0$ and $V_z\ne 0;V_x =0$ give the same probability distribution of zero-energy modes. This is due to the fact that the two terms are mapped onto each other ($V_x\to V_z$ and $V_z\to-V_x$) by the unitary transformation $(\sigma_0+i\sigma_y)/\sqrt{2}$. This is no longer the case for a 2D system consisting of coupled nanowires with different superconducting phases, where we find that the direction of the Zeeman field affects the number and the probability distribution of gapless modes.

\subsection{Two-wire system}

\begin{figure}[tb]
 \includegraphics[width=\columnwidth]{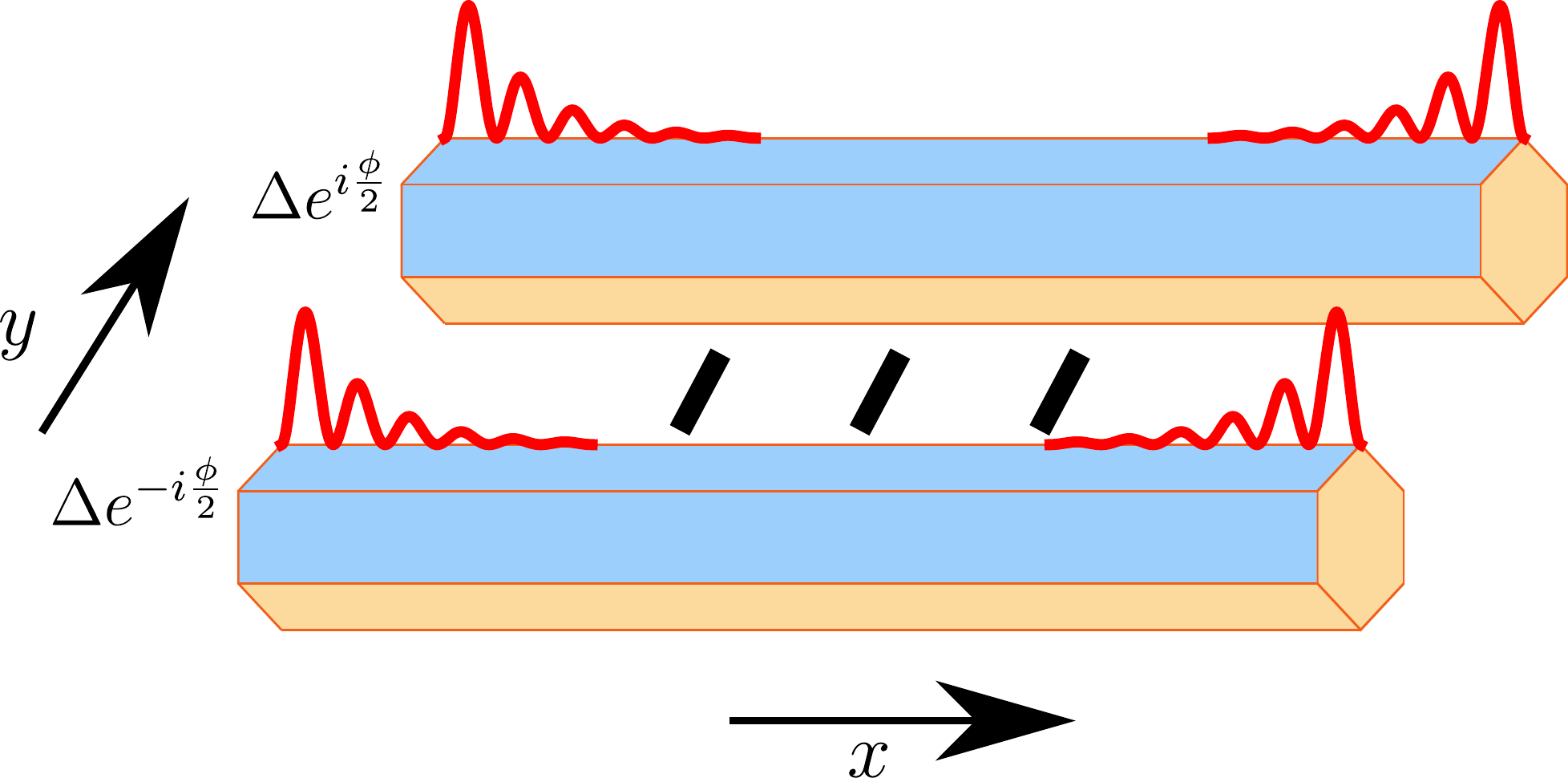}
  \caption{System composed of two coupled wires with opposite superconducting phases. Each nanowire (yellow) is covered by a superconductor (blue), and hosts a pair of MBS (red). The thick black lines indicate the coupling between the wires.}
\label{fig:2wires_system}
\end{figure}

We now proceed with the model of two Majorana nanowires with opposite superconducting phases (Fig.~\ref{fig:2wires_system}) describing how the energy of the mid-gap states can be tuned by changing the phase difference.\cite{Haim2019}

The Hamiltonian density takes the form
\begin{align} \label{eq:2wires_ham}
\begin{split}
\mathcal{H} (k_x) = &  [2t_x(1- \cos{k_x})  + 2t_y -\mu] \eta_0 \tau_z \sigma_0  + \\ 
&   V_x \eta_0 \tau_0 \sigma_x  +  V_z \eta_0 \tau_0 \sigma_z + \\
& \Delta \cos{\frac{\phi}{2}} \; \eta_0 \tau_x \sigma_0  +  \Delta \sin{\frac{\phi}{2}} \; \eta_z \tau_y \sigma_0 +  \\
& \alpha \sin{k_x} \eta_0\tau_z \sigma_y  - 2 t_y \eta_x \tau_z \sigma_0 + \beta  \eta_y \tau_z \sigma_x,
\end{split}
\end{align}
where $t_y$ is a spin-conserving hopping term between wires and $\beta$ is a term which describes a hopping which flips the electron spin, a consequence of the Rashba SOC in the y-direction. The superconducting phase difference between the wires is $\phi$ and the Pauli matrix $\eta$ denotes the wire space, while all other symbols retain their meaning. Particle-hole symmetry is preserved with the same operator, $\mathcal{P} = \tau_y \sigma_y \mathcal{K}$, and we can define a new chiral symmetry operator, $\mathcal{C} = \eta_x \tau_y \sigma_y$, and, consequently, $\mathcal{T} = \eta_x \mathcal{K}$.

From now on, unless otherwise specified, the parameters used in the calculations are $\mu = 0$, $t_x=1.7$, $\alpha = 2.5$, $\Delta = 2.5$, $t_y =0.4$ and $\beta = 0.8$.  Furthermore, we will consider two separate cases: $V_{x}= 4, V_z = 0$ or $V_{z}= 4, V_x = 0$. 

\begin{figure}[tb]
 \includegraphics[width=\columnwidth]{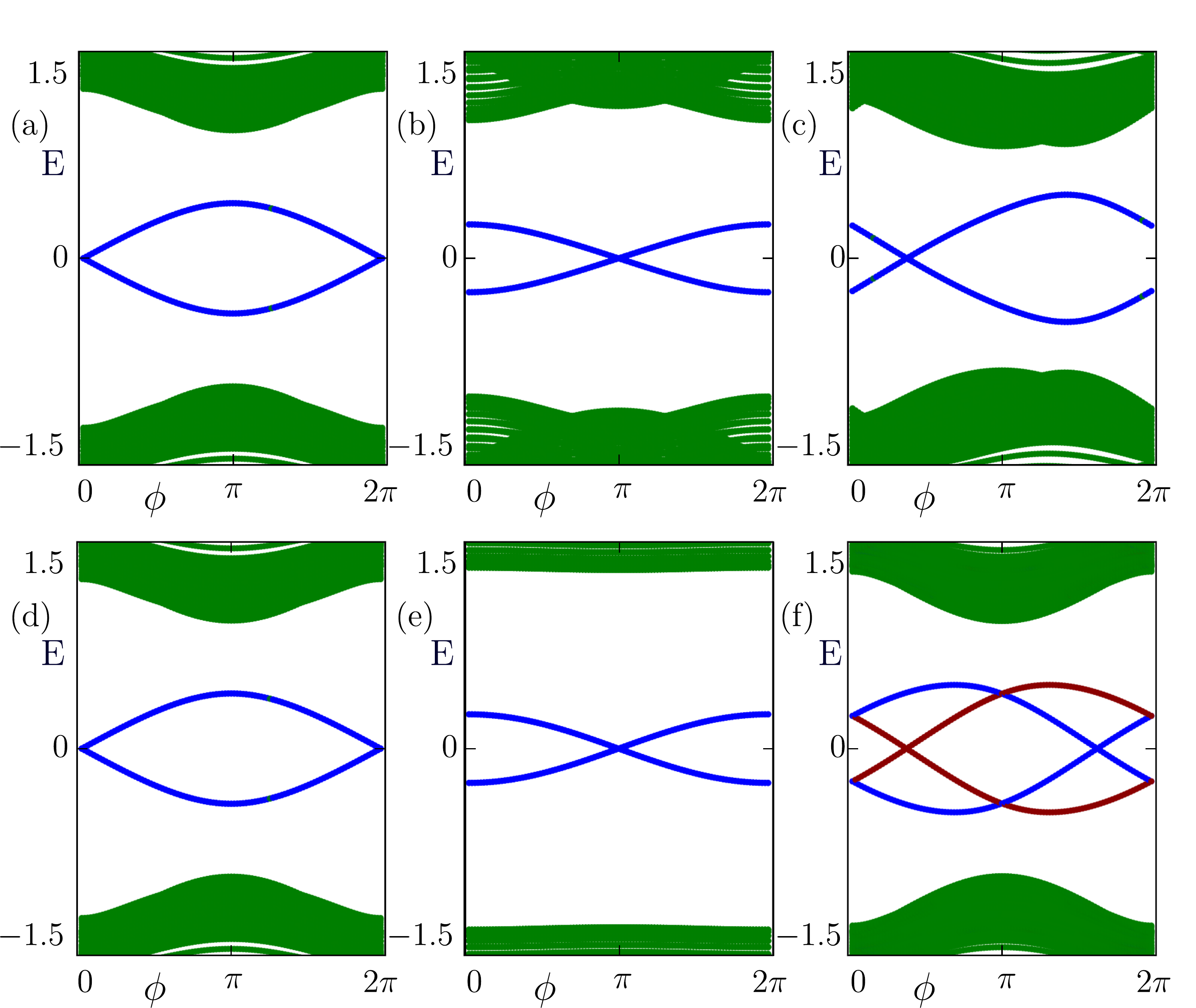}
  \caption{Spectrum of the two-wire system as a function of the phase difference between wires. We color the bulk modes in green and edge modes in blue (and red, if the mirror symmetry is broken). In (a-c), the Zeeman field points in the x-direction; (a) $t_y \ne 0$ and $\beta  = 0$,  (b) $t_y = 0$ and $\beta  \ne 0$,  (c) $t_y \ne 0$ and $\beta  \ne 0$; (d-f) The Zeeman field points in the z-direction;  (d) $t_y \ne 0$ and $\beta  = 0$,  (e) $t_y = 0$ and $\beta  \ne 0$,  (f) $t_y \ne 0$ and $\beta  \ne 0$. In all cases, we consider two wires that are $80$ sites long. For panels (a-e), the mirror-symmetric boundary modes are represented with blue, whereas for panel (f), this symmetry is broken and the Majorana modes localized on the right/left boundary of the system are shown in red/blue, respectively.}
\label{fig:2wires_spectrum}
\end{figure}

Throughout the following, the key insight we will use to generate a SOTSC is that, by altering the phase difference between the wires, for instance by applying a supercurrent, it is possible to tune the splitting between their MBS. This is shown in Fig.~\ref{fig:2wires_spectrum}, where the spectrum of the two-wire system is plotted as a function of $\phi$.\footnote{All numerical simulations are performed using Kwant,\cite{Groth2014} and the code is included as part of the Supplementary Material.} The splitting depends both on the type of inter-wire coupling (only $t_y$, only $\beta$, or both of them), as well as on the direction of the Zeeman field.

If $V_z=0$ and $V_x\neq 0$ (top panels of Fig.~\ref{fig:2wires_spectrum}), the interaction strength between the pair of MBS on the left and right side of the system is identical. This is due to a mirror symmetry, ${\cal M}_x=\sigma_x$, such that $\sigma_x H(-k_x) \sigma_x = H(k_x)$. When only the spin-conserving hopping is present, the phase dependence of the mid-gap energy levels takes the form $\pm\sin{\frac{\phi}{2}}$, and the two levels cross at $\phi=0$ (Fig.~\ref{fig:2wires_spectrum}a). On the other hand, when only the spin-flip hopping is included, the crossing occurs at $\phi=\pi$, with end-state energies behaving as $\pm \cos{\frac{\phi}{2}}$ (Fig.~\ref{fig:2wires_spectrum}b). These numerical results are confirmed in Appendix \ref{app:pert} using first order perturbation theory in $t_y$ and $\beta$. When both coupling terms are nonzero, the crossing point will happen at a generic value of $\phi$, which depends on $t_y$ and $\beta$ (Fig.~\ref{fig:2wires_spectrum}c). In this case, by tuning the phase difference, the MBS coupling can be made either stronger or weaker than if the wires were proximitized by the same superconductor (meaning $\phi=0$).

Setting $V_x=0$ and $V_z\neq0$ results in similar MBS couplings (bottom panels of Fig.~\ref{fig:2wires_spectrum}). The principal difference between the effect of $V_z$ and $V_x$ is the fact that, for $V_z\neq0$, there is no unique mirror symmetry operator in the x-direction. Instead, one can define two different $\mathcal{M}_x$ for each of the couplings $t_y$ ($\mathcal{M}_x = \sigma_z$, Fig.~\ref{fig:2wires_spectrum}d) and $\beta$ ($\mathcal{M}_x = \eta_z \sigma_z$, Fig.~\ref{fig:2wires_spectrum}e). As such, a broken mirror-symmetry phase is realized once both couplings are present, one in which the MBS splitting on the right boundary of the system is different from that on the left boundary. This is shown in Fig.~\ref{fig:2wires_spectrum}f, where the pair of end modes on the right and left is shown in red and blue, respectively. Due to the symmetry $\eta_x\sigma_z H(k_x,\phi) \eta_x\sigma_z = H(-k_x, -\phi)$, left-end modes at phase difference $\phi$ have the same energy as right-end modes at $-\phi$.

\subsection{Many-wires system}

\begin{figure*}[t]
 \includegraphics[width=\textwidth]{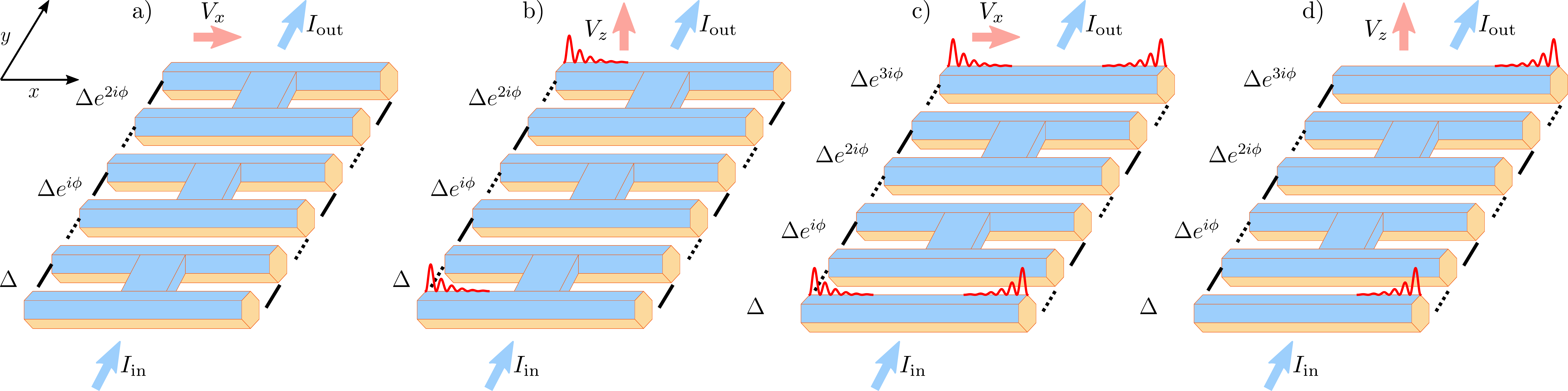}
  \caption{Array of nanowires realizing a SOTSC. Panels (a) and (b) show the model given by the Hamiltonian Eqs.~\eqref{eq:2D_ham} and \eqref{eq:conf1}, whereas panels (c) and (d) show the setup corresponding to Eqs.~\eqref{eq:2D_ham} and \eqref{eq:conf2}. The Zeeman field points in the x-direction for panels (a) and (c), whereas it points in the z-direction for panels (b) and (d). In all cases, superconducting bridges (blue) connect pairs of wires, leading to a vanishing phase difference between them. Phase-biasing the device such that unconnected superconductors have a small, positive phase difference (for instance, by means of a supercurrent $I_{\rm in/out}$) will then only alter the MBS coupling between disconnected pairs of wires. The resulting dimerization pattern (solid and dashed lines) drives the system into topologically different phases. These include a trivial phase (a), a nontrivial SOTSC with four Majorana corner modes (red, panel c), as well as SOTSC phases with two corner modes, on the left (b) or right (d) edge. Reversing the direction of the supercurrent (blue arrows) causes topological phase transitions, changing the corner mode distribution from that of panel (a) to that of panel (c), and from panel (b) to panel (d), respectively. Similarly, rotating the direction of the Zeeman field amounts to interchanging (a) with (b) and (c) with (d).}
\label{fig:2DSystem}
\end{figure*}

Having shown how superconducting phase differences can be used to engineer end-state energies, we now use this control parameter to generate a SOTSC phase. The setup, consisting of an array of equally spaced nanowires, is shown in Fig.~\ref{fig:2DSystem}. We introduce superconducting bridges which connect neighboring wires into different patterns, corresponding to panels a, b and panels c, d of Fig.~\ref{fig:2DSystem}. We assume that these superconducting ``connectors'' are placed far from the ends of the wires, such that they do not alter the MBS wavefunctions. However, due to these bridges, pairs of nanowires are effectively proximitized by the same superconductor, leading to a vanishing phase difference between them. Disconnected pairs of wires, on the other hand, may still show a nonzero $\phi$, for instance when a supercurrent is passed through the device.

The real-space Hamiltonian written in the basis $\Psi_{x,y}^{\dagger} = ( \psi_{x,y, \uparrow}^{\dagger} , \psi_{x,y, \downarrow}^{\dagger}, \psi_{x,y, \downarrow}, - \psi_{x,y, \uparrow})$, where the subscripts $x,y$ denote the positions of sites in the tight-binding model, reads
\begin{align}\label{eq:2D_ham}
\begin{split}
H_{\rm TB} & = \sum_{x,y} \Psi_{x,y}^{\dagger} [(2t_x  + 2t_y -\mu) \tau_z \sigma_0 + V_x  \tau_0 \sigma_x  + \\
& V_z \tau_0 \sigma_z + \Delta \cos{\phi(y)}  \tau_x \sigma_0  +
 \Delta \sin{\phi(y)} \tau_y \sigma_0  ] \Psi_{x,y}  \\
& + \{ \Psi_{x,y}^{\dagger} [-t_x \tau_z \sigma_0 + i \frac{\alpha}{2} \tau_z \sigma_y ] \Psi_{x + 1,y} + \rm{h.c.} \} \\
& + \{ \Psi_{x,y}^{\dagger} [-t_y \tau_z \sigma_0 + i \frac{\beta}{2} \tau_z \sigma_x ] \Psi_{x,y + 1} + \rm{h.c.} \}, 
\end{split}
\end{align}
where $\phi(y)$ is the superconducting phase that depends on the coordinate $y$. For the setup shown in Fig.~\ref{fig:2DSystem}a, b it takes values

\begin{equation}\label{eq:conf1}
\phi(y)=\left\{\
                \begin{array}{ll}
                   0 $ , $ y = 1, 2\\
                   \phi $ , $ y = 3,4\\
                   2 \phi $ , $ y = 5,6  \\
                   \vdots \\
                   (\frac{N_{\rm wires}}{2} - 1)\phi $ , $ y = N_{\rm wires} -1, N_{\rm wires}.
               \end{array}
              \right.
\end{equation}
where, $N_{\rm wires}$ is the (even) number of Majorana nanowires in the y-direction. The other possibility, Fig.~\ref{fig:2DSystem}c, d, has
\begin{equation}\label{eq:conf2}
\phi(y)=\left\{\
                \begin{array}{ll}
                  0 $ , $ y = 1\\
                   \phi $ , $ y = 2,3\\
                   2  \phi $ , $ y = 4,5  \\
                   \vdots \\
                   \frac{N_{\rm wires}}{2} \phi $ , $ y = N_{\rm wires}.
               \end{array}
              \right.
\end{equation}
In both arrangements, the superconducting phase difference between adjacent superconductors is the same throughout the system and equals $\phi$. 

We first discuss the case when only the Zeeman field in the x-direction is present. Given current nanowire fabrication techniques, we expect this to be the most relevant scenario, since the superconducting shell covering the wires is typically very thin, such that a magnetic field perpendicular to the wires could destroy superconductivity.

For the arrangement of nanowires depicted in Fig.~\ref{fig:2DSystem}a, the SOTSC phase is obtained when the phase difference is tuned such that the energy of MBS that originate from different superconductors is increased compared to the case $\phi = 0$. On the contrary, the setup of Fig.~\ref{fig:2DSystem}c requires a stronger coupling between the MBS of wires proximitized by the same superconductor. This produces a dimerization of the MBS couplings in adjacent wires, such that the edge perpendicular to the wires (the $y$-edge) becomes a nontrivial Kitaev chain.\cite{Kitaev2001} The result is a phase with a gapped bulk, gapped edges, but Majorana modes localized at the corners. The latter occur due to the fact that both edges are topologically nontrivial: the edge parallel to $x$ is the last topological nanowire of the array, whereas the edge parallel to $y$ is a nontrivial Kitaev chain.

\begin{figure}[tb]
 \includegraphics[width=\columnwidth]{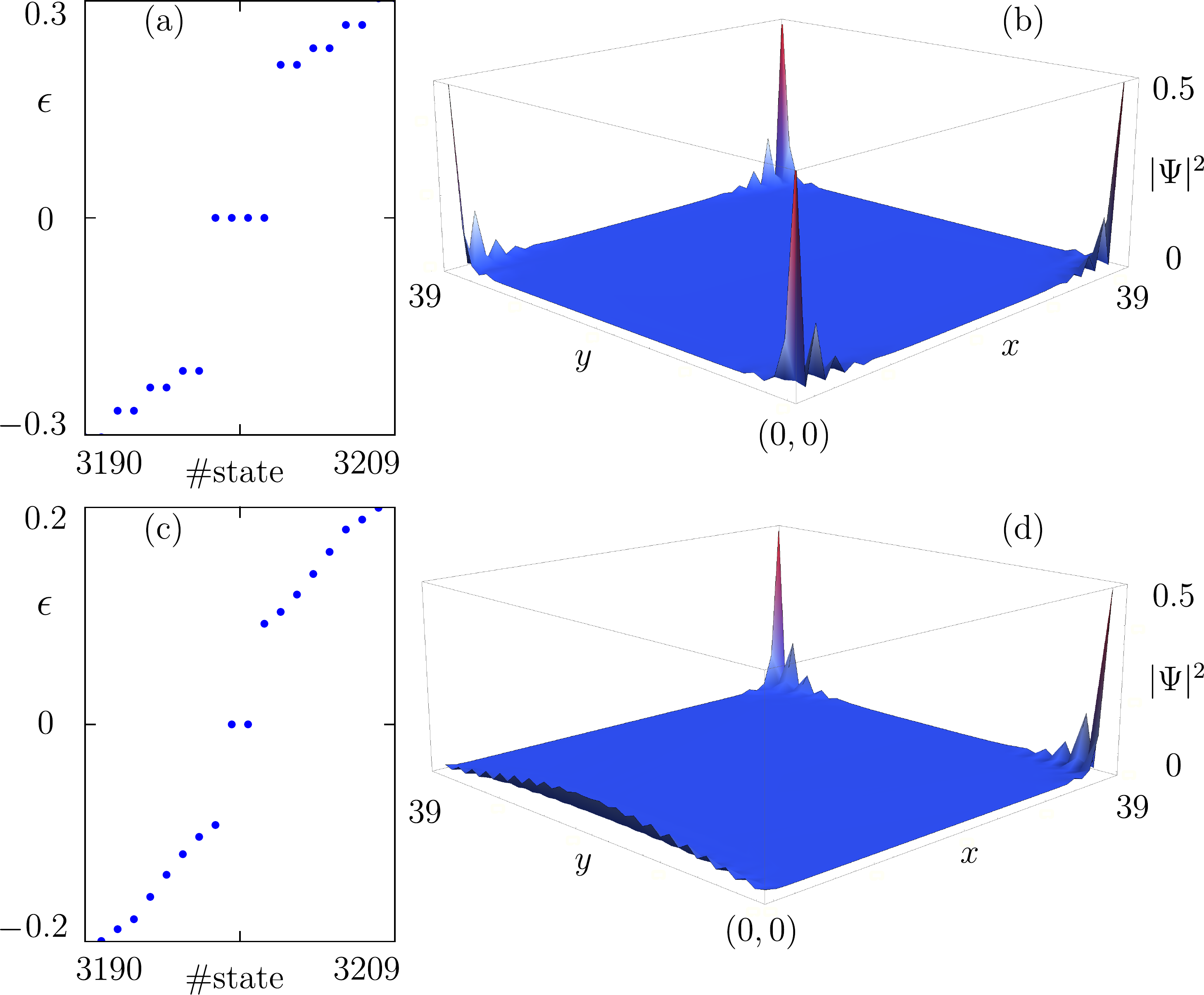}
  \caption{(a,b)/(c,d) The spectrum and the spatial distribution of four/two gapless modes for a system consisting of $40 \times 40$ sites, with the Zeeman field in the x-/z-direction. In both cases, the phase difference is $\phi = 4.5$. }
\label{fig:spect_prob}
\end{figure}

Given the spectrum shown in Fig.~\ref{fig:2wires_spectrum}c, the appropriate range to observe the nontrivial phase in the setup of Fig.~\ref{fig:2DSystem}a is $\phi\in(\sim 2.26,2 \pi)$. We choose a value $\phi = 4.5$ and calculate the energy spectrum and the probability distribution of zero-energy modes (see Figs.~\ref{fig:spect_prob}a and \ref{fig:spect_prob}b). As the mirror symmetry in the x-direction is not broken, we expect four zero-energy modes that are pinned to the corners, which is confirmed by our calculations. For the other arrangement of nanowires (Fig.~\ref{fig:2DSystem}c), one should tune the phase difference in the regime where the energy of MBS decreases compared to $\phi = 0$ to obtain the SOTSC phase.

For the sake of completeness, we also examine the case of a Zeeman field pointing in the z-direction. Note that an externally applied magnetic field which is perpendicular to the nanowires will lead to a complicated phase relation across the Josephson junctions, in addition to the afore mentioned possibility of destroying superconductivity altogether. It is however possible to generate such a Zeeman term in the absence of external fields, by taking advantage of the ferromagnetic ordering of atoms deposited on a superconductor.\cite{Nadj-Perge2013,Nadj-Perge2014}
In this case, the behavior of the Josephson junctions may be treated in the same way as before. 

For the system in which $V_z\neq0$, Fig.~\ref{fig:2wires_spectrum}f suggests that any nonzero $\phi$ produces a SOTSC phase with only two gapless corner modes, regardless of a particular realization of the system. This is due to the fact that MBS on the left and right edges of the array will be dimerized in an opposite fashion once a small phase difference is applied.
For instance, setting $ \phi = 4.5$, the configuration given by Eq.~\eqref{eq:conf1}, supports two zero-energy modes (Fig.~\ref{fig:spect_prob}c) located on two right corners (Fig.~\ref{fig:spect_prob}d). The same phase difference for the other configuration would produce two MBS located on the left corners, as this realization requires smaller energy splittings compared to their value at $\phi = 0$ (Fig.~\ref{fig:2wires_spectrum}f). The phase difference $\phi = \pi$ is special, however, as there are four corner modes for a configuration as in Fig.~\ref{fig:2DSystem}b and no corner modes for the setup of Fig.~\ref{fig:2DSystem}d.  

\section{Characterization} \label{sec:characterization}

In order to show the topological nature of the setup shown in Fig.~\ref{fig:2DSystem}, we calculate the topological invariant using scattering theory.\cite{Akhmerov2011, Fulga2011, Fulga2012, Geier2018} This invariant is based on the parity of the number of MBS at one corner and can be used to characterize both first-order \cite{Akhmerov2011, Fulga2011, Fulga2012} and higher-order phases.\cite{Geier2018} 

To determine the scattering matrix, a unitary matrix that connects incoming to outgoing modes, we consider a four-terminal transport geometry, consisting of a 2D system, given by the Hamiltonian Eq.~\eqref{eq:2D_ham}, and four normal-metal leads oriented along the x-direction, each attached to one corner. Each lead $l$ supports four incoming modes $\Psi^{\rm in}_l$ and four outgoing mode $\Psi^{\rm out}_l$ at the Fermi level, corresponding to spin and particle-hole degrees of freedom.

In this setup, the scattering matrix $S$ has a $4 \times 4$ block structure, whose diagonal blocks are reflection matrices $r_l$, $l=1,...,4$, and whose non-diagonal blocks are transmission matrices $t_{ln}$ ($l,n = 1,..,4$ and $l \ne n$) from one lead to another.
In class D,\cite{Altland1997} with the particle-hole symmetry $\mathcal{P} = \tau_y \sigma_y \mathcal{K}$, the scattering matrix obeys 
\begin{equation}\label{eq:constraint}
S = \tau_y \sigma_y S^* \tau_y \sigma_y.
\end{equation}
The procedure to obtain this relation is detailed in Appendix \ref{app:smat}. 
Given the relation Eq.~\eqref{eq:constraint}, it follows that the determinant of the full scattering matrix, $\det S$, as well as that of each reflection block $\det r_l$ are real. This allows to define a topological invariant: the sign of the determinant of the reflection matrix\cite{Akhmerov2011,Geier2018}
\begin{equation}\label{eq:top_inv}
Q_l = \rm{sign} (\det{ r_l}),
\end{equation}
In total, four invariants can be defined, one for each corner, and they are not mutually independent, as MBS always appear in pairs.

The reflection matrix is a matrix block of the form
\begin{equation}\label{refl}
r_l =
  \begin{pmatrix}
   r^{ee}_l & {r^{eh}_l}  \\
    {r^{he}_l} & {r^{hh}_l} 
  \end{pmatrix},
\end{equation}
where $r^{ee}$ and $r^{hh}$ are $2 \times 2$ matrices that represent normal reflections (scatterings of electrons into electrons and holes into holes), whilSzczecine matrices $r^{eh}$ and $r^{he}$ describe Andreev reflection processes.

Once the bulk and the edges of a 2D system are gapped, the incoming mode can only be reflected. Then, reflection matrices $r_l$ become unitary due to current conservation and their determinants satisfy $|\det{ r_l}| = 1$. If the incoming mode is reflected on a Majorana bound state, $\det{ r_l} = -1$ due to a $\pi$ phase shift in the scattered states.\cite{Akhmerov2011} This contrasts the case of no Majorana corner states, for which $\det{ r_l} = 1$. 

The phase transition occurs once the y-edge gap is closed and two counter-propagating edge states appear, protected by the translation symmetry.
At this point, the topological invariant has to change sign $(Q = 0)$, which is possible as the reflection matrix is not unitary anymore. This change of sign is accompanied by the appearance of the quantized value of thermal conductance, defined as $G = G_{\rm th} \rm{tr}$$(t^{\dagger} t)$.\cite{Akhmerov2011} Here, $t$ is the transmission matrix between the two relevant leads and $G_{\rm th} = \frac{\pi^2  k_B^2 T}{6h}$ is the thermal conductance quantum. Note that this phase transition point separating the trivial and SOTSC phase is precisely a weak-TSC, hosting gapless Majorana modes on the y-edges, modes which are protected by translation symmetry.

\begin{figure}[tb]
 \includegraphics[width=\columnwidth]{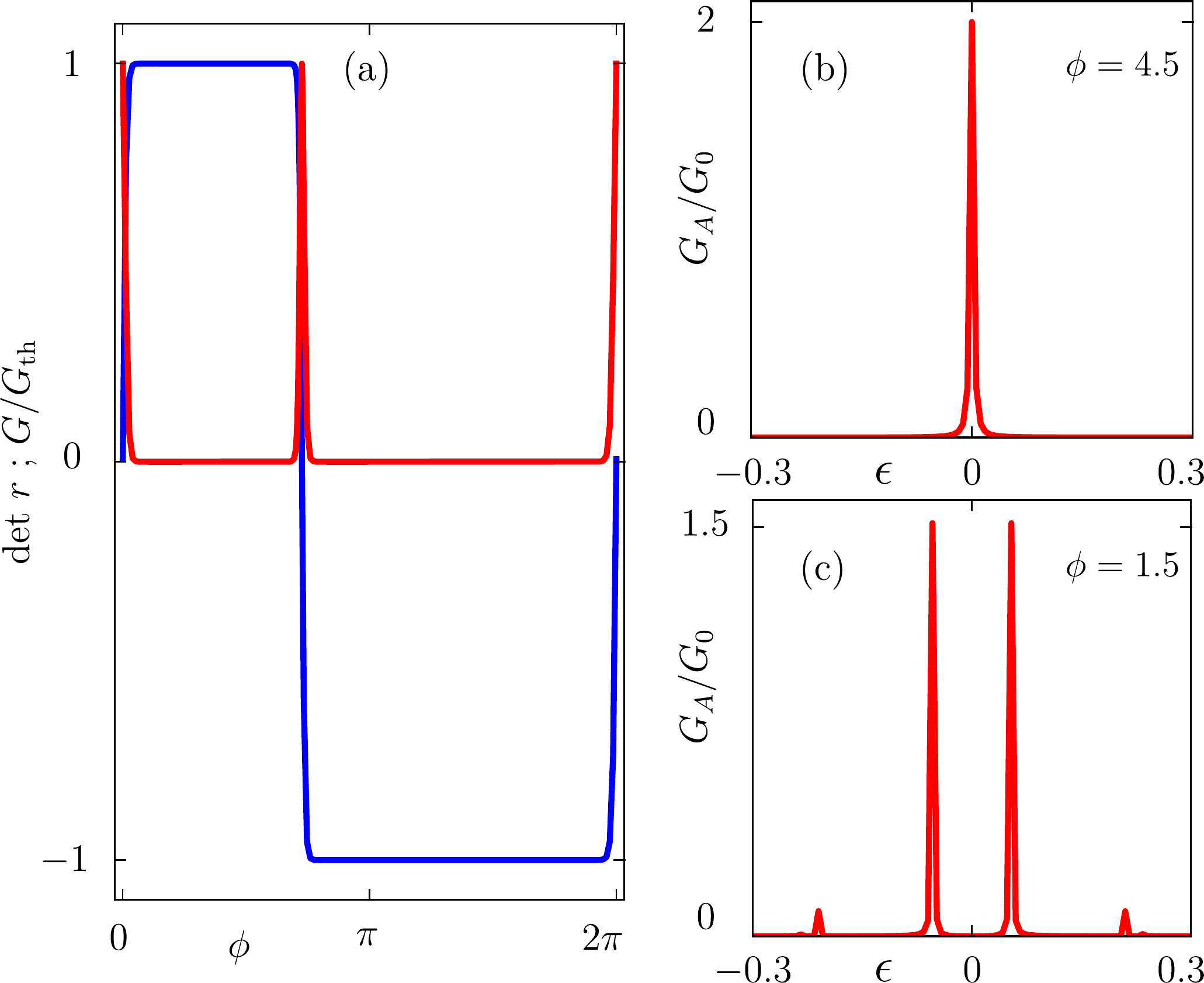}
  \caption{(a) The topological invariant (blue) and the thermal conductance (red) as functions of the superconducting phase difference, for a system size of $80 \times 80$ sites. (b) The Andreev conductance in a SOTSC phase, once $\phi =4.5$. (c) The Andreev conductance in a trivial phase, once $\phi = 1.5$. All calculations are done for the Zeeman field in the x-direction. }
\label{fig:inv_conductance_Vx}
\end{figure}

In Fig.~\ref{fig:inv_conductance_Vx}a, we show the value of $\det{r}$ (blue line) as the phase difference between the nanowires is changed for a system, represented in Fig.~\ref{fig:2DSystem}a, with the Zeeman field in the x-direction. Here, we have dropped the index $l$ as all gapless modes have the same dependence on $\phi$. Furthermore, we use $\det{ r}$ as it is a continuous function whose values match $Q$ far away from the phase transition. The boundaries of the topological phase agree with the values of $\phi= 0$ and $\phi \simeq 2.26$ that produce uniform energy splittings of MBS throughout the system (see Fig.~\ref{fig:2wires_spectrum}c). These transitions, are accompanied by quantized thermal conductance peaks, represented with red lines in Fig.~\ref{fig:inv_conductance_Vx}a.

We further study the nature of the corner modes by calculating the Andreev conductance, one of the main experimental tools to confirm the presence of MBS.\cite{Fu2009,Akhmerov2009,Mourik2012,Deng2016,Zhang2018a} The latter lead to resonant Andreev reflection, giving rise to a quantized value of the Andreev conductance $G_A = 2 G_0$ according to the relation\cite{Blonder1982}
\begin{equation} \label{eq:andreev_cond}
G_A = G_0 [N - \tr{(r_{ee}^{}r_{ee}^{\dagger})} + \tr{(r_{eh}^{}r_{eh}^{\dagger})}],
\end{equation}
where $G_0 = \frac{e^2}{h}$, $\tr$ denotes the trace, and $N$ is the number of incoming electron modes in the lead ($N=2$ in our case). 
The numerical results of Fig.~\ref{fig:inv_conductance_Vx}b show that in the SOTSC phase there is a quantized zero bias conductance peak, whereas this peak is absent in Fig.~\ref{fig:inv_conductance_Vx}c, which represents a trivial phase.

\begin{figure}[tb]
 \includegraphics[width=\columnwidth]{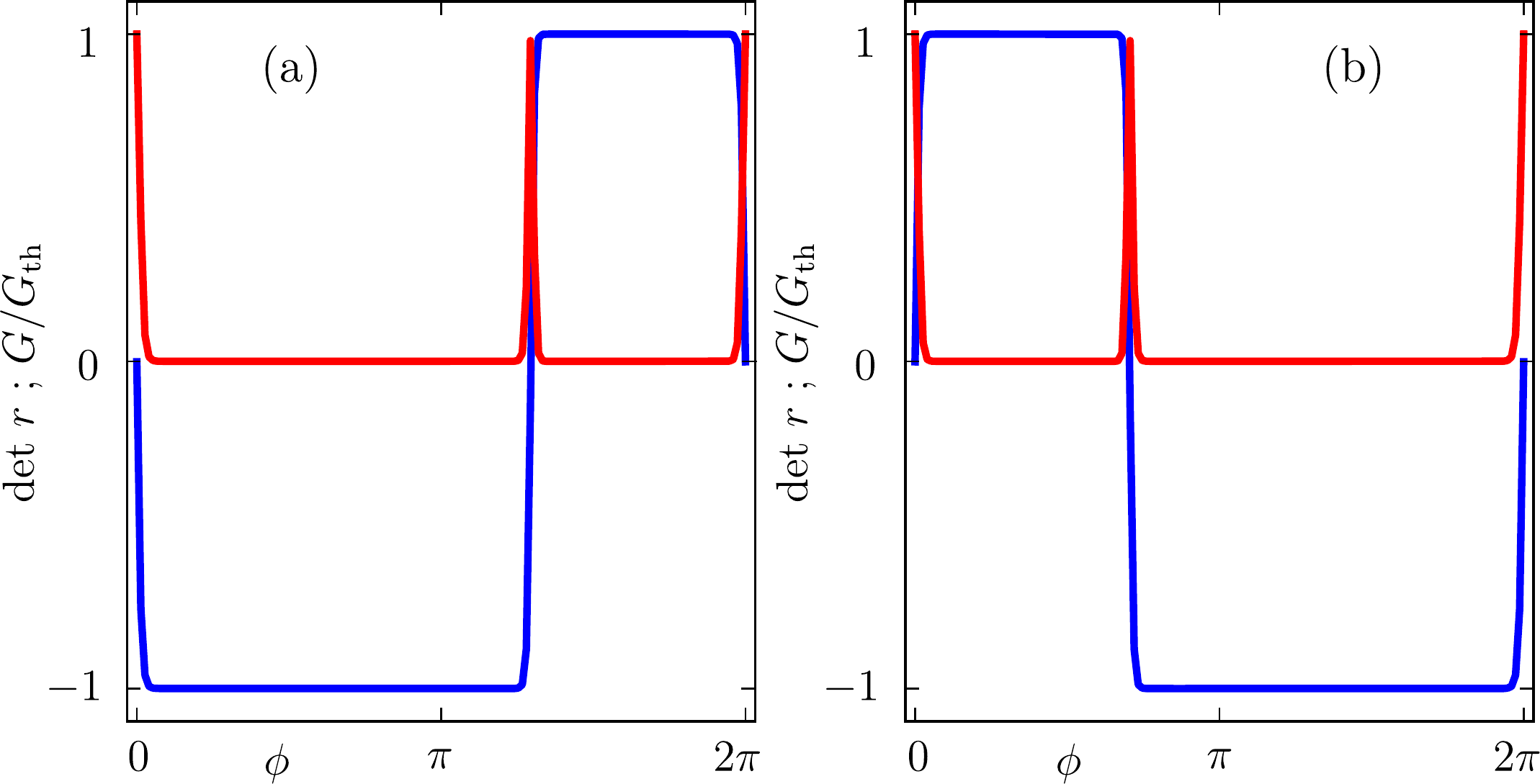}
  \caption{ The topological invariant (blue) and the thermal conductance (red) as functions of the superconducting phase, for the system size $80 \times 80$ sites and $V_z\neq0$. Panels (a)/(b) are obtained using left/right leads.}
\label{fig:inv_conductance_Vz}
\end{figure}

We now discuss the case of the Zeeman field in the z-direction and for the configuration given in Fig.~\ref{fig:2DSystem}b. We expect that the leads attached on left/right corners show different dependence of the topological invariant and the thermal conductance on $\phi$. We plot the topological invariant/thermal conductance calculated using left leads in Fig.~\ref{fig:inv_conductance_Vz}a and the same quantities for right leads are given in Fig.~\ref{fig:inv_conductance_Vz}b. On the left side of the system, the phase transition occurs for $\phi = 0$ and $\phi \simeq 4.02$. On the right side of the system, the nontrivial phase happens when $\phi \in (\sim 2.26, 2 \pi)$. As expected for this arrangement of wires, all four leads measure a nontrivial phase at $\phi = \pi$. 

\section{Disorder} \label{sec:disorder}

\begin{figure*}[tb]
 \includegraphics[width=\textwidth]{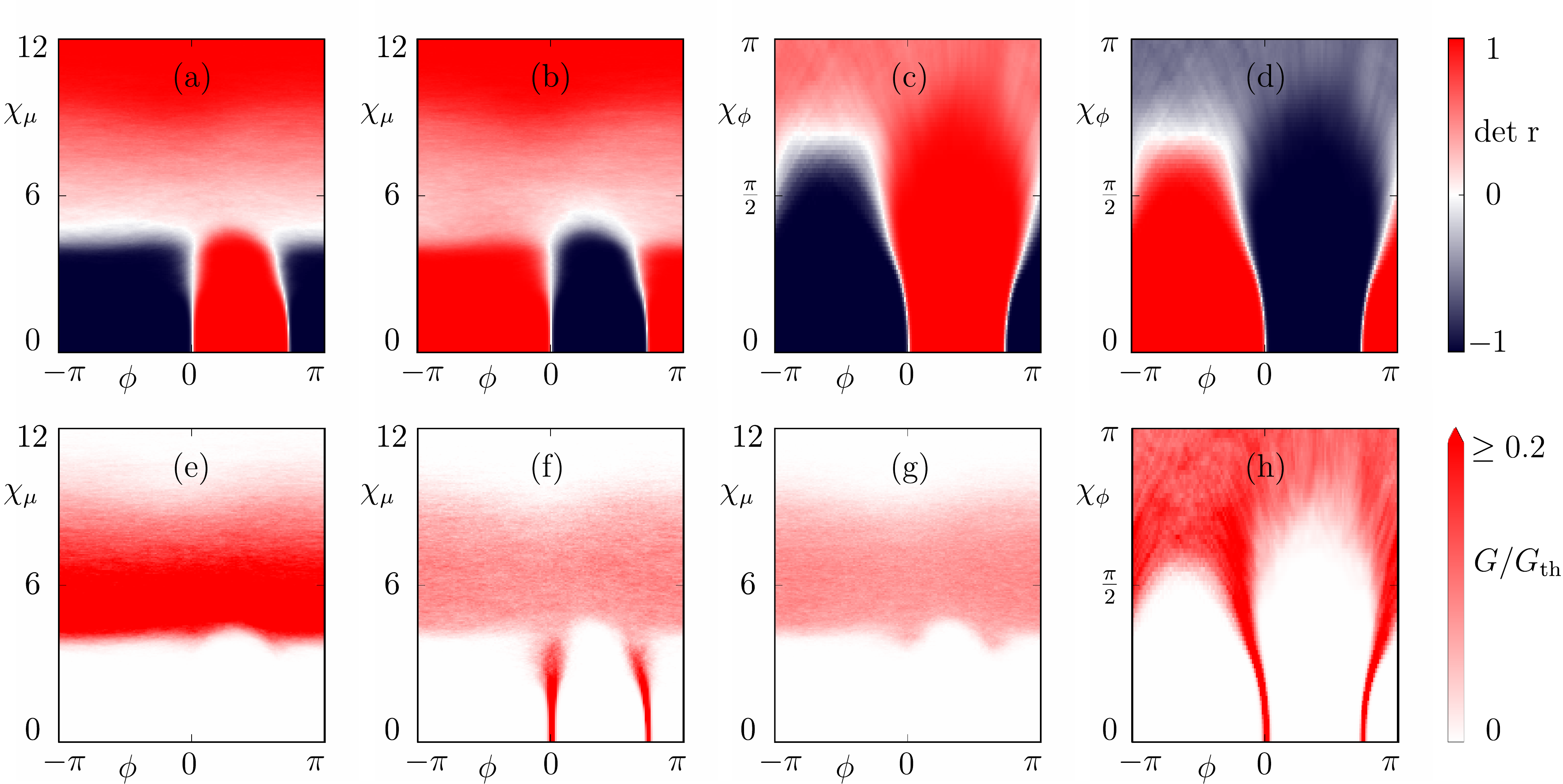}
  \caption{(a-d) Phase diagrams, obtained as a function of phase difference (horizontal axis) and disorder strength (vertical axis), either in the chemical potential $\mu$ (a,b) or in the phases $\phi$ (c,d). In the top panels, the color scale indicates the topological invariant, whereas in the bottom panels it denotes the thermal conductance. Panels (a) and (c) correspond to the system described by Eqs.~\eqref{eq:2D_ham} and \eqref{eq:conf1}, while (b) and (d) to Eqs.~\eqref{eq:2D_ham} and \eqref{eq:conf2}. These phase diagrams are obtained by averaging topological invariants obtained from 301 disorder realizations.
(e-g) The averaged value of x-edge/y-edge/bulk conductance, respectively, for the disorder in the chemical potential $\mu$. (h) The averaged y-edge conductance, corresponding to panels (c,d). In all calculations, the system size is $80 \times 80$ sites.}
\label{fig:disorder}
\end{figure*}

To model realistic systems, we have to take into account the effects of disorder. Although the breaking of the translational invariance does not affect the validity of our topological invariant,\cite{Akhmerov2011} it makes invariants at the corners mutually independent. Various kinds of disorder act differently on the topological phase in Majorana nanowires.\cite{Lutchyn2010,Stanescu2011,Akhmerov2011,Brouwer2011,Adagideli2014} In particular, it was found that MBS are robust against short-ranged disorder in chemical potential in single-band nanowires\cite{Adagideli2014} and their multi-band generalizations.\cite{Lutchyn2010,Stanescu2011} In the following, we will study the effect of chemical potential fluctuations in the nanowires on the SOTSC phase. Furthermore, we anticipate that the distance between the superconductors that form Josephson junctions will not be uniform throughout the realistic system. Such imperfections create fluctuations of the critical currents $I_{C, j}$ ($j$ denotes the junction),\cite{A-M, Abrikosov} which in turn lead to fluctuations of the phase differences acquired across the junctions, once an external supercurrent $I$ is imposed. Therefore, on the Hamiltonian level, we simulate this type of disorder by randomizing the phase differences of neighboring superconductors. 

Disorder in the chemical potential can originate from impurities in the nanowire. In our simulations, we model these fluctuations by replacing $\mu\to \mu+v_\mu$, where $v_{\mu}$ is drawn independently for each lattice site from the uniform distribution $[-\chi_{\mu},\chi_{\mu}]$, with $\chi_\mu$ the strength of disorder. The phase diagram, as a function of the phase $\phi$ and disorder strength, is represented in Fig.~\ref{fig:disorder}a for the arrangement of nanowires given in Fig.~\ref{fig:2DSystem}a and in Fig.~\ref{fig:disorder}b for the arrangement of nanowires as in Fig.~\ref{fig:2DSystem}c. In both cases, we consider a SOTSC phase with four gapless modes in the clean limit, while the case of a non-vanishing $V_z$ will be addressed in Appendix \ref{app:disorder}. The diagram represents an averaged value of four topological invariants, each calculated for one corner, over $301$ independent disorder realizations. These results are accompanied by the calculations of x-edge/y-edge/bulk conductance, represented in Figs. \ref{fig:disorder}e, \ref{fig:disorder}f and \ref{fig:disorder}g, respectively. The x-edge/y-edge conductance is calculated using transmission matrices between the leads that share the same horizontal/vertical edges, while the bulk conductance is calculated using a system with periodic boundary conditions in one direction. The phase boundaries look the same for the two different system configurations, so we plot only once the dependence of these conductances on $\phi$ and disorder strength.

We notice that the topological phase is stable up to moderate strengths ($0<\chi_{\mu}<4$) of disorder. Furthermore, as smaller values of $\chi_{\mu}$ do not strongly break the translational invariance in the y-direction, the thermal conductance along the y-edges remains close to unity (bright red line in Fig.~\ref{fig:disorder}f). Stronger disorder leads to a metallic phase, characterized by the appearance of x-edge and bulk conductance.\cite{Medvedyeva2010,Fulga2012a} Finally, the system is in the trivial Anderson insulator phase for very large disorder strengths ($\chi_{\mu}>8$), with $Q= 1$ for any value of $\phi$. In Fig.~\ref{fig:disorder}a, we notice the appearance of the topological Anderson insulator phase,\cite{Li2009,Groth2009,Song2012,Lv2013,Adagideli2014} as the boundaries of the SOTSC phase around $\phi = 2.26$ are extended compared to the clean limit. In effect, this means that a system which it topologically trivial in the clean limit can be rendered nontrivial once disorder is introduced.

As already pointed out, experimental realizations of our system might have different phase differences between the adjacent superconductors. Therefore, we also model disorder in the phases of superconducting pairing. Phase diagrams of setups shown in Figs.~\ref{fig:2DSystem}a and \ref{fig:2DSystem}c are presented in Figs.~\ref{fig:disorder}c and \ref{fig:disorder}d, respectively. More precisely, we take $\phi_{\rm{dis}}(y) = \phi(y) + v_{\phi}$, where $v_{\phi}$ is uniformly distributed in the range $[-\chi_{\phi},\chi_{\phi}]$, such that $\braket{\phi_{\rm{dis}}(y)} = \phi(y)$. As before, we consider $301$ independent disorder realizations.
For the configuration proposed in Fig.~\ref{fig:2DSystem}a, we observe that disorder in $\phi$ reduces the extent of the topological phase in $\phi$ as compared to the clean limit. The opposite happens for the arrangement of nanowires like in Fig.~\ref{fig:2DSystem}c, which represents another example of a higher-order topological Anderson insulator. As the two configurations share the same phase boundaries, the y-edge conductance, represented in Fig.~\ref{fig:disorder}h, has the same behavior in both cases. For an almost-clean system, this conductance will be close to one, while larger fluctuations in the phases diminish its value. On the other hand, phase disorder does not cause any x-edge or bulk conductance, since for our parameters the bulk states remain gapped for any $\phi$ (see Fig.~\ref{fig:2wires_spectrum}c). We see that these configurations have different tendencies for large disorder strength, i.e. the SOTSC phase either increases or decreases. However, in limit of small phase deviations, the boundary between the two phases remains close to the one in the clean limit, such that a small value of $\phi$ is still sufficient to cause a topological phase transition.

\section{Conclusion} \label{sec:conclusion}

Higher-order topological phases present an opportunity to explore novel phenomena.
In this work, we constructed a 2D model of a second-order topological superconductor controlled by means of superconducting phase differences. The latter lead to dimerized MBS couplings in an array of Majorana nanowires, allowing to access both the SOTSC phase as well as its transition to a trivial phase. Phase-biasing the device may be achieved by means of a supercurrent (as shown in Fig.~\ref{fig:2DSystem}), or alternatively by using flux loops.

From an experimental point of view, our model offers several advantages. First, it only requires conventional, s-wave superconductors, as well as the fabrication techniques recently developed in the context of Majorana nanowires,\cite{Krogstrup2015,Gazibegovic2017,Casparis2018,Vaitiekenas2018a} including the deposition of a superconductor on a nanowire using \textit{in situ} methods. Second, it allows for a topological phase to be reached for a range of phase differences, without the latter having to be set to a particular value. In fact, an arbitrarily small value of $\phi$ is sufficient to dimerize the MBS couplings of the nanowire array and reach a topological phase, though it may be that corner modes have a very large localization length in this case. Furthermore, if the SOTSC phase is controlled by a Josephson current, the latter could be made time-dependent, opening the possibility to study periodically-driven SOTSC phases.\cite{Bomantara2018,Rodriguez-Vega2018,Huang2018}

As the nontrivial phase is protected only by particle-hole symmetry, our model is robust to fluctuations in both the chemical potential and phase differences between the nanowires. Depending on the configuration of the system, we have shown that the topological phase can expand in the presence of disorder in the chemical potential, leading to a second-order topological Anderson phase. Furthermore, simulations reveal that randomness of the superconducting phases has a similar effect. Thus, depending on the way we couple the superconductors, the topological phase can either enlarge or decrease.  

\section{Acknowledgments} \label{sec:acknowledgement}
We thank Ulrike Nitzsche for technical assistance. This work was supported by the DFG through the W{\"u}rzburg-Dresden Cluster of Excellence on Complexity and Topology in Quantum Matter -- \textit{ct.qmat} (EXC 2147, project-id 39085490).

\appendix

\section{Corrections to energies of gapless modes} \label{app:pert}

For very small inter-wire coupling, the correction to the energies of the non-perturbed Hamiltonian can be found using perturbation theory. Throughout this Section, we closely follow the derivation presented in Ref.~\onlinecite{Klinovaja2012}. This procedure requires knowing the initial wave-functions, and for this reason we start from a system of two uncoupled nanowires with opposite superconducting phases.
Then, the BdG Hamiltonian $H = \int dx \Psi^{\dagger} \mathcal{H} \Psi$ is defined through the Hamiltonian density
\begin{align} \label{eq:MFHam_Vz}
\begin{split}
& \mathcal{H} =  [\frac{ - \hbar^2 \partial_x^2}{2m} -\mu] \eta_0 \tau_z \sigma_0  +  V_z \eta_0 \tau_0 \sigma_z + \\ 
& \Delta \cos{\frac{ \phi}{2}} \eta_0 \tau_x \sigma_0  + \Delta \sin{\frac{\phi}{2}} \eta_z \tau_y \sigma_0  - i  \alpha \partial_x \eta_0 \tau_z \sigma_y, 
\end{split}
\end{align} 
in the basis 

$\Psi^{\dagger} = (\psi_{1,\uparrow}^{\dagger}, \psi_{1,\downarrow}^{\dagger}, \psi_{1,\downarrow}^{}, -\psi_{1,\uparrow}^{}, \psi_{2, \uparrow}^{\dagger}, \psi_{2,\downarrow}^{\dagger}, \psi_{2,\downarrow}^{}, -\psi_{2,\uparrow}^{})$. Here, $\psi^\dag_{j,\sigma}$ is a fermionic creation operator on wire $j$, with spin $\sigma$.

Finding zero-energy solutions of the above Hamiltonian density can be simplified as it is a block-diagonal matrix. In fact, it is sufficient to solve a $4 \times 4$ matrix differential equation with real pairing strength and then restore phases $\exp(\pm i \phi/2)$ into the solution [$\Psi_{\phi} (x) = U \Psi_{\phi=0} (x)$] of  Eq.~\eqref{eq:MFHam_Vz} with the unitary transformation $U_{\phi} = \exp{ \left(-\frac{i}{2} \frac{\phi}{2} \eta_z \tau_z \sigma_0 \right) }$.\cite{Jiang2011} Thus, the relevant Hamiltonian density is
\begin{align} \label{eq:mnanowire_ham}
\begin{split}
 \mathcal{H}_{\rm{SNW}} =   [\frac{- \hbar^2 \partial_x^2}{2m} -\mu] \tau_z  +  V_z  \sigma_z +  
 \Delta \tau_x - i \alpha \partial_x  \tau_z \sigma_y, 
\end{split}
\end{align} 
in the basis $\Psi^{\dagger} = (\psi_{\uparrow}^{\dagger}, \psi_{\downarrow}^{\dagger}, \psi_{\downarrow}^{}, -\psi_{\uparrow}^{})$ and we have dropped the wire index. 
We will work in the limit where the SOC energy is the largest energy scale of the problem and for simplicity, we assume $\mu = 0$. As we use Kwant\cite{Groth2014} for numerical calculations, where hard-wall boundary conditions are assumed, our analytical solution in semi-infinite geometry has to obey $\Psi(x = 0) = 0$ and $\Psi (x = \infty) \rightarrow 0$. Thus, we concentrate on the zero-energy state localized around $x=0$. The use of semi-infinite geometry to calculate this wave-function is a valid approximation as MBS at two ends of the system should have a negligible overlap in order to avoid hybridization.\cite{Klinovaja2012}

Now, we perform a unitary transformation $U_{\rm{rot}} \mathcal{H}_{\rm{SNW}} U_{\rm{rot}}^{\dagger}$ where $U_{\rm{rot}} = \tau_0 \frac{\sigma_0 - i \sigma_x}{\sqrt{2}}$ to bring the SOC term into a diagonal form
\begin{align} \label{eq:mnanowire_ham_rotated}
\begin{split}
 \mathcal{H}'_{\rm{SNW}} =   \frac{ - \hbar^2\partial_x^2}{2m} \tau_z  -  V_z  \sigma_y +  
 \Delta \tau_x - i \alpha \partial_x  \tau_z \sigma_z, 
\end{split}
\end{align} 
and
\begin{align*} 
\begin{split}
 \Psi^{'\dagger} & = \frac{1}{\sqrt{2}} (\psi_{\uparrow}^{\dagger} + i \psi_{\downarrow}^{\dagger}, \psi_{\downarrow}^{\dagger} + i \psi_{\uparrow}^{\dagger}, \psi_{\downarrow}^{}- i \psi_{\uparrow}^{}, -\psi_{\uparrow}^{} + i \psi_{\downarrow}^{}) \\
& \equiv  (\psi_{\uparrow}^{'\dagger}, \psi_{\downarrow}^{'\dagger}, \psi^{'}_{\downarrow}, -\psi^{'}_{\uparrow}).
\end{split}
\end{align*}

The eigenenergies of the system are obtained using the Fourier transformation and are $E^2_{\pm} = \frac{\hbar^4}{4m^2} (k_x^4 + 4 k_{\rm{SO}}^2 k_x^2) + \Delta^2 + V_z^2 \pm 2\sqrt{\Delta^2 V_z^2 +  \frac{\hbar^4 k_x^4}{4m^2} (V_z^2 +  \frac{\hbar^4} {m^2} k_{\rm{SO}}^2 k_x^2)}$, where $k_{\rm{SO}} = \frac{m \alpha}{\hbar^2}$ and $k_x$ denotes the momentum. The gap $E_{\rm{gap}} = 2 E_{-}$ is reduced for two values of $k_x$; at $k_x=0$ this gap equals $\Delta - V_z$ and its sign determines whether the system is in the topological phase $(V_z > \Delta)$ or not, while the gap at $k_x = k_F = 2 k_{\rm{SO}}$ is always nonzero as it is induced by superconductivity. Thus, all states close to $k_x=0$ form the interior branch while the states close to $k_x= k_F$ belong to the exterior branch of the spectrum.

In the regime of strong SOC, the magnetic field and proximity-induced superconductivity are considered as perturbations. To transfer the problem in the rotating frame, we use a spin-dependent gauge transformation\cite{Klinovaja2012,Braunecker2015} $[\psi'_{\sigma} (x) = \exp(-i \sigma k_{\rm{SO}} x) \tilde{\psi}'_{\sigma} (x)]$, where $\sigma = \pm 1$ denotes spin-up/spin-down. 

Before the gauge transformation, kinetic energy and SOC of the electrons reads
\begin{equation*}
\mathcal{H}'_{\rm{T+SOC}} =  -\frac{\hbar^2 \partial_x^2}{2m}\sigma_0 - i \alpha \partial_x \sigma_z,
\end{equation*}
and after the transformation 
\begin{equation*}
\tilde{H}^{'}_{\rm{T+SOC}} = \sum_{\sigma,\sigma'} \int dx \; \tilde{\psi}^{'\dagger}_{\sigma} (x) e^{i \sigma k_{\rm{SO}} x} \mathcal{H}'_{\rm{T+SOC}} e^{-i \sigma' k_{\rm{SO}} x} \tilde{\psi}^{'}_{\sigma'} (x),
\end{equation*}
it becomes
\begin{equation*}
\mathcal{\tilde{H}}'_{\rm{T+SOC}} =  -\frac{\hbar^2}{2m} (\partial_x^2  + k_{\rm{SO}}^2),
\end{equation*}
as the first-order derivatives in $x$ cancel exactly. Thus, this transformation eliminates the SOC term and now, the spectrum of $\tilde{\mathcal{H}}^{'}_{\rm{T+SOC}}$ consists of two parabolas centered at $k_x = 0$ $[E = \frac{\hbar^2}{2m}(k_x^2 -  k_{\rm{SO}}^2)]$. Then, around the Fermi points $k = \pm k_{\rm SO}$, the spectrum can be linearized, i.e. the electron operator becomes
\begin{equation}\label{eq:el_operators}
\tilde{\psi}^{'}_{\sigma} (x) = \tilde{R}^{'}_{\sigma} e^{i k_{\rm{SO}} x} +\tilde{L}^{'}_{\sigma} e^{-i k_{\rm{SO}} x}, 
\end{equation}
where $\tilde{R}^{'}_{\sigma}/\tilde{L}^{'}_{\sigma}$ denote right/left movers.

In the linearized system, we ignore the second derivatives of $\tilde{R}^{'}_{\sigma}/\tilde{L}^{'}_{\sigma}$. Further, in the rotating frame, we also neglect the fast-oscillating terms $[\exp(\pm i n  k_{\rm{SO}} x)$, where $n>1]$. 
Then, the term $\tilde{H}^{'}_{\rm{T+SOC}}$ becomes
\begin{align} \label{eq:lin_momentum}
\begin{split}
\tilde{H}^{'}_{\rm{T+SOC}} = - i \alpha \int dx \; & [ \tilde{R}^{'\dagger}_{\uparrow} \partial_x \tilde{R}^{'}_{\uparrow} + 
\tilde{R}^{'\dagger}_{\downarrow} \partial_x \tilde{R}^{'}_{\downarrow}  \\
& - \tilde{L}^{'\dagger}_{\uparrow} \partial_x \tilde{L}^{'}_{\uparrow} - 
\tilde{L}^{'\dagger}_{\downarrow} \partial_x \tilde{L}^{'}_{\downarrow}]
\end{split}
\end{align}

The spin-dependent gauge transformation also impacts the Zeeman field term, as it becomes a helical field that rotates in the plane that is perpendicular to the spin-orbit coupling vector in order to minimize the energy.\cite{Braunecker2015} The form of this field $\mathbf{V}_z = V_z(\hat{y} \cos{2 k_{\rm SO} x} + \hat{x} \sin{2 k_{\rm SO} x})$, where $\hat{x}$/$\hat{y}$ are unit vectors in the x-/y-direction, is inherited from the Zeeman Hamiltonian  
\begin{align} \label{eq:zeeman_ham}
\begin{split}
\tilde{H}'_z &= -V_z \sum_{\sigma,\sigma'} \int dx \; \tilde{\psi}_{\sigma}^{'\dagger} e^{i \sigma  k_{\rm{SO}} x} (\sigma_y)_{\sigma,\sigma'} e^{-i \sigma'  k_{\rm{SO}} x} \tilde{\psi}^{'}_{\sigma'} \\
& = -V_z \int dx \; [-i \tilde{\psi}_{\uparrow}^{'\dagger} e^{2 i k_{\rm{SO}} x} \tilde{\psi}^{'}_{\downarrow} + i \tilde{\psi}_{\downarrow}^{'\dagger} e^{-2i  k_{\rm{SO}} x} \tilde{\psi}^{'}_{\uparrow}] \\
&=  -V_z \int dx \; [-i \tilde{R}_{\uparrow}^{' \dagger} \tilde{L}^{'}_{\downarrow} + i \tilde{L}_{\downarrow}^{' \dagger} \tilde{R}_{\uparrow}^{'}],
\end{split}
\end{align}
where we have dropped the fast oscillating terms to obtain the last row.

The initial superconducting Hamiltonian was 
\begin{equation*}
H^{'}_{\rm{SC}} = -\Delta \int dx \; [{\psi}_{\uparrow}^{'}{\psi}_{\downarrow}^{'} - {\psi}_{\downarrow}^{'} {\psi}_{\uparrow}^{'}  + \rm{h.c.}] ,
\end{equation*}
and after the transformation, it becomes 
\begin{align} \label{eq:sc_ham}
\begin{split}
\tilde{H}^{'}_{\rm{SC}} =  -\Delta \int dx \; & [\tilde{R}_{\uparrow}^{'}\tilde{L}_{\downarrow}^{'} + \tilde{L}_{\uparrow}^{'}\tilde{R}_{\downarrow}^{'} - \\ 
& \tilde{R}_{\downarrow}^{'} \tilde{L}_{\uparrow}^{'} - \tilde{L}_{\downarrow}^{'} \tilde{R}_{\uparrow}^{'}  + \rm{h.c.}]. 
\end{split}
\end{align}

Finally, we define two basis vectors $\phi^{'i \dagger} = (\tilde{R}_{\uparrow}^{'\dagger}, \tilde{L}_{\downarrow}^{'\dagger}, \tilde{L}_{\downarrow}^{'},- \tilde{R}_{\uparrow}^{'})$ and  $\phi^{'e \dagger} = (\tilde{L}_{\uparrow}^{'\dagger}, \tilde{R}_{\downarrow}^{'\dagger}, \tilde{R}_{\downarrow}^{'},- \tilde{L}_{\uparrow}^{'})$. Then, we are able to construct Hamiltonians corresponding to interior ($k \sim 0$) and exterior ($k \sim 2 k_{\rm{SO}}$) branches in the laboratory frame from terms in Eqs.~\eqref{eq:lin_momentum},~\eqref{eq:zeeman_ham} and~\eqref{eq:sc_ham}:
\begin{equation*}
\tilde{H}^{' l} = \frac{1}{2} \int dx (\psi^{' l})^{ \dagger} \mathcal{\tilde{H}}^{' l} \psi^{' l},
\end{equation*}
where $l = i,e$ and
\begin{align}\label{eq:branches_hams_rotated}
\begin{split}
& \mathcal{\tilde{H}}^{' i} = - i \alpha \tau_z \sigma_z \partial_x + \Delta \tau_x - V_z \sigma_y \\
& \mathcal{\tilde{H}}^{' e} =  i \alpha \tau_z \sigma_z \partial_x + \Delta \tau_x.
\end{split}
\end{align}
The unitary transformation $U_{\rm{rot}}$ applied on Hamiltonians in Eq.~\eqref{eq:branches_hams_rotated} rotates the SOC/Zeeman energy into the form as in Eq.~\eqref{eq:mnanowire_ham}. Thus, these Hamiltonians, in the basis $\phi^{i \dagger} = (\tilde{R}_{\uparrow}^{\dagger}, \tilde{L}_{\downarrow}^{\dagger}, \tilde{L}_{\downarrow},- \tilde{R}_{\uparrow})/\phi^{e \dagger} = (\tilde{L}_{\uparrow}^{\dagger}, \tilde{R}_{\downarrow}^{\dagger}, \tilde{R}_{\downarrow},- \tilde{L}_{\uparrow})$, become
\begin{align} \label{eq:branches_hams}
\begin{split}
& \mathcal{\tilde{H}}^{i} = - i \alpha \tau_z \sigma_y \partial_x + \Delta \tau_x + V_z \sigma_z \\
& \mathcal{\tilde{H}}^ e =  i \alpha \tau_z \sigma_y \partial_x + \Delta \tau_x
\end{split}
\end{align}
Zero-energy solutions obey $H^l \Psi^l = 0$ and to solve this set of equations, we assume the ansatz $\Psi^l = \exp(\kappa_l x) \psi^l$, where $\psi^l$ is a spinor. Then, the eigenenergies are 
\begin{align} \label{eq:branches_energies}
\begin{split}
& (\tilde{E}^{i})^2 = - \alpha^2 \kappa^2_i + (\Delta \pm V_z)^2 \\
& (\tilde{E}^ e)^2 =  - \alpha^2 \kappa^2_e + \Delta^2.
\end{split}
\end{align}
 
For zero-energy modes, we obtain $\kappa_i = \pm \frac{\Delta \pm V_z}{\alpha}$ and $\kappa_e = \pm \frac{\Delta}{\alpha}$. As we look for a normalizable solution in the segment $[0,\infty)$, this leaves us with $\kappa_{i,1} = \frac{\Delta - V_z}{\alpha}$,$\kappa_{i,2} = - \frac{\Delta + V_z}{\alpha}$ as well as  $\kappa_e = - \frac{\Delta}{\alpha}$.
Then, to obtain these eigenfunctions in the basis $\tilde{\Psi}^{\dagger} = (\tilde{\psi}_{\uparrow}^{\dagger}, \tilde{\psi}_{\downarrow}^{\dagger}, \tilde{\psi}_{\downarrow},- \tilde{\psi}_{\uparrow})$, we add phases $\exp{(\pm i k_{\rm{SO}} x})$ associated with right/left movers to spinor parts of the ansatz $\Psi^l$. Finally, we obtain four eigenfunctions 
\begin{align*}
 \tilde{\Psi}_1^i (x) &= \frac{C_{i,1}}{2}\begin{pmatrix}
  -e^{-i k_{\rm{SO}} x } \\
   e^{i k_{\rm{SO}} x } \\
e^{-i k_{\rm{SO}} x } \\
e^{i k_{\rm{SO}} x }
\end{pmatrix} e^{\frac{\Delta-V_z}{\alpha}x}, 
\end{align*}
\begin{align*}
 \tilde{\Psi}_2^i (x) &  = \frac{C_{i,2}}{2}\begin{pmatrix}
  i e^{-i k_{\rm{SO}} x }  \\
    - i e^{i k_{\rm{SO}} x } \\
 i e^{-i k_{\rm{SO}} x } \\
 i e^{i k_{\rm{SO}} x } 
\end{pmatrix}  e^{-\frac{\Delta+V_z}{\alpha}x},
\end{align*}
\begin{align*}
 \tilde{\Psi}_1^e (x) &= \frac{C_e}{2}\begin{pmatrix}
  -e^{i k_{\rm{SO}} x } \\
   e^{-i k_{\rm{SO}} x } \\
e^{i k_{\rm{SO}} x } \\
e^{-i k_{\rm{SO}} x }
\end{pmatrix} e^{-\frac{\Delta}{\alpha}x}, \; \rm{and}
\end{align*}
\begin{align*}
 \tilde{\Psi}_2^e (x) &  = \frac{C_e}{2}\begin{pmatrix}
 - e^{i k_{\rm{SO}} x }  \\
    -e^{-i k_{\rm{SO}} x } \\
 -e^{i k_{\rm{SO}} x } \\
 e^{-i k_{\rm{SO}} x } 
\end{pmatrix}  e^{-\frac{\Delta}{\alpha}x}.
\end{align*}
Here, coefficients $C_{i,1},C_{i,2},C_e$ are normalization coefficients that should be obtained from $\int_0^{\infty} dx |\Psi^l(x)|^2 = 1$. 

\begin{figure}[tb]
 \includegraphics[width=\columnwidth]{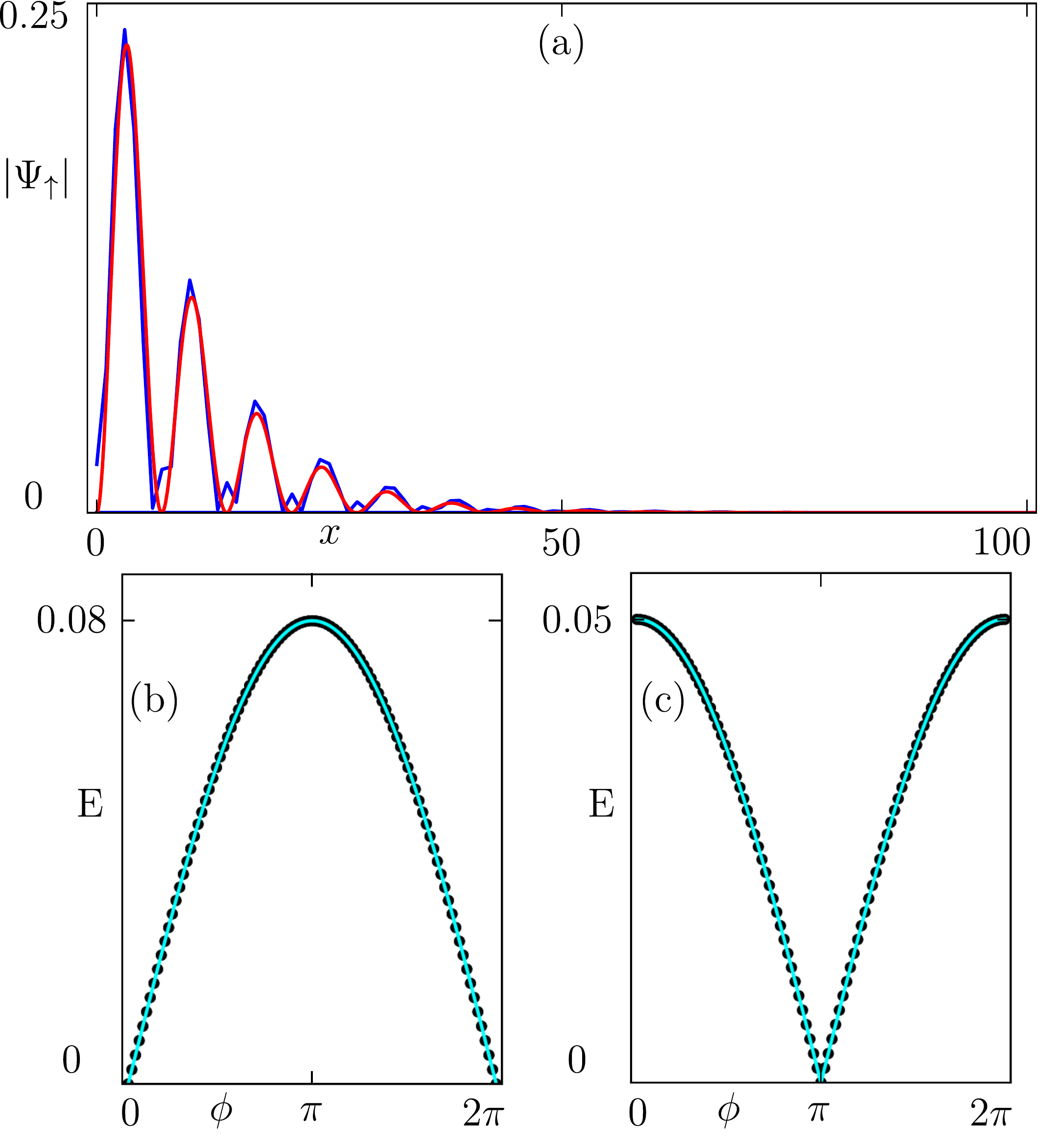}
  \caption{(a) To compare analytics and numerics, we plot the absolute value of the first component of a Majorana wave-function localized around $x = 0$. The analytical solution [Eq.~\eqref{eq:MF_wavefunc}] is represented by a red line while the numerical wave-function, calculated using Kwant, is given in blue. For the numerical solution, we consider an $800$ sites long wire with the lattice constant $a=1$ and the hopping strength $t_x=10$. Other parameters are $m = -1/20, \mu = 0, \alpha = 9, \Delta = 1, \phi = 0$ and $V_z = 2$. (b) For $t_y = 0.08$ and $\beta=0$, the analytical solution $ t_y \sin(\phi/2)$ given with a blue line matches a numerical calculation, represented with black points. (c) For $\beta = 0.08$ and $t_y=0$, the analytical solution $ \beta \Delta \cos(\phi/2)/V_z$ plotted in blue matches a numerical simulation, represented with black points. Intra-wire parameters used in (b) and (c) are $\mu =0, t_x = 1.7, \alpha = 2.5, \Delta = 2.5$ and $V_z = 4$. }
\label{fig:analytics_numerics}
\end{figure}

The wave-function of a Majorana bound state should be a linear superposition of these four eigenfunctions. However, upon imposing the boundary condition $\tilde{\Psi} (x = 0) = 0$, we see that the only possible combination is $\tilde{\Psi} = \tilde{\Psi}_1^i - \tilde{\Psi}_1^e$. By using the relation $\Psi_{\sigma} (x) = \exp(-i \sigma k_{\rm{SO}} x) \tilde{\Psi}_{\sigma} (x)$, we obtain the wave-function in the laboratory frame
\begin{align} \label{eq:MF_wavefunc}
 \Psi (x) &= \frac{C}{2}\begin{pmatrix}
  -1 \\
  1 \\
1 \\
1
\end{pmatrix} e^{\frac{\Delta-V_z}{\alpha}x} - 
\frac{C}{2}\begin{pmatrix}
  -e^{2 i k_{\rm{SO}} x} \\
   e^{-2 i k_{\rm{SO}} x} \\
e^{2 i k_{\rm{SO}} x} \\
e^{-2 i k_{\rm{SO}} x}
\end{pmatrix} e^{-\frac{\Delta}{\alpha}x},
\end{align}
where $C = \sqrt{\frac{2  \Delta (V_z - \Delta) (V_z^2 +4 \frac{m^2\alpha^4}{\hbar^4})}{ V_z \alpha [(V_z -2 \Delta)^2 + 4 \frac{m^2\alpha^4}{\hbar^4}]}}$. We find a good overlap of the analytical solution with the numerical one (see Fig.~\ref{fig:analytics_numerics}a) in the limit where SOC energy is the largest energy scale. 

The wave-functions that are solutions of Eq.~\eqref{eq:MFHam_Vz} for zero energy [$\Psi_1 (x)/\Psi_2 (x)$] are obtained after applying a unitary transformation $U_{\phi} = \exp{ \left(- \frac{i}{2} \frac{\phi}{2} \eta_z \tau_z \sigma_0 \right) }$ to $(1 \; 0)^{\dagger} \otimes \Psi^{\dagger} (x) / (0 \; 1)^{\dagger} \otimes \Psi^{\dagger} (x)$ and are equal to
\begin{align*}
 \Psi_1 (x) &= \frac{C}{2}\begin{pmatrix}
  -e^{-\frac{i \phi}{4}} \\
   e^{-\frac{i \phi}{4}} \\
e^{\frac{i \phi}{4}} \\
e^{\frac{i \phi}{4}}\\
0\\
0\\
0\\
0
\end{pmatrix} e^{\frac{\Delta-V_z}{\alpha}x} - 
\frac{C}{2}\begin{pmatrix}
  -e^{-\frac{i \phi}{4} +2 i k_{\rm{SO}} x} \\
   e^{-\frac{i \phi}{4} - 2 i k_{\rm{SO}} x} \\
e^{ \frac{i \phi}{4} + 2 i k_{\rm{SO}} x} \\
e^{\frac{i \phi}{4} -2 i k_{\rm{SO}} x} \\
0\\
0\\
0\\
0
\end{pmatrix} e^{-\frac{\Delta}{\alpha}x}
\end{align*}
and
\begin{align*}
 \Psi_2 (x) &= \frac{C}{2}\begin{pmatrix}
0 \\
0 \\
0 \\
0 \\
  -e^{\frac{i \phi}{4}} \\
   e^{\frac{i \phi}{4}} \\
e^{-\frac{i \phi}{4}} \\
e^{-\frac{i \phi}{4}}
\end{pmatrix} e^{\frac{\Delta-V_z}{\alpha}x} - 
\frac{C}{2}\begin{pmatrix}
0 \\
0 \\
0 \\
0 \\
  -e^{\frac{i \phi}{4} +2 i k_{\rm{SO}} x} \\
   e^{\frac{i \phi}{4} -2 i k_{\rm{SO}} x} \\
e^{ -\frac{i \phi}{4} + 2 i k_{\rm{SO}} x} \\
e^{-\frac{i \phi}{4} -2  i k_{\rm{SO}} x} 
\end{pmatrix} e^{-\frac{\Delta}{\alpha}x}.
\end{align*}

Having two zero-energy states requires a degenerate perturbation theory to calculate the corrections to energy of these modes once the inter-wire coupling is present. Thus, we look at the matrix
\begin{equation} \label{eq:pert_theory}
H_{\rm correction}=
 \left( {
\begin{array}{cc}
\bra{\Psi_{1}} H_{\rm coupling}  \ket{ \Psi_{1}} & \bra{\Psi_{1}}H_{\rm coupling}  \ket{ \Psi_{2}} \\
\bra{\Psi_{2}} H_{\rm coupling}  \ket{ \Psi_{1}} & \bra{\Psi_{2}} H_{\rm coupling}  \ket{ \Psi_{2}}
\end{array} } \right  ),
\end{equation}
where $H_{\rm coupling}$ denotes the coupling between the nanowires. 

We extract the coupling terms from Eq.~\eqref{eq:2wires_ham}, i.e. $H_{t_y} =  - t_y \eta_x \tau_z $ and $H_{\beta} =  \beta \eta_y \tau_z \sigma_x $. Thus, only the off-diagonal terms in Eq.~\eqref{eq:pert_theory} are non-zero and complex-conjugate to each other.

For the normal hopping, one of these off-diagonal terms is 
\begin{align*}
\begin{split}
&\bra{\Psi_{1}}H_{\rm coupling}  \ket{ \Psi_{2}} =   C^2 t_y (e^{i\phi/2}-e^{-i \phi/2}) \times \\
& \int_0^{\infty} e^{-\frac{V_z}{ \alpha} x}[\cos{(\frac{2m\alpha}{\hbar^2} x)} - \cosh{(\frac{V_z - 2\Delta}{\alpha} x)}] dx \\
& = - \frac{ t_y (e^{i\phi/2}-e^{-i \phi/2})}{2},
\end{split}
\end{align*}
therefore, giving the matrix 
\begin{equation*} 
H_{t_y}=
 \frac{t_y}{2} \left( {
\begin{array}{cc}
0  & - (e^{i \phi/2} -e^{-i \phi/2} ) \\
 (e^{-i \phi}-e^{-i \phi/2}) & 0
\end{array} } \right  ),
\end{equation*}
whose eigenvalues are $\pm t_y \sin{(\frac{\phi}{2})}$, which is also confirmed by a numerical simulation presented in Fig.~\ref{fig:analytics_numerics}b. 

Once only the spin-orbit coupling in the y-direction is present, the off-diagonal term becomes  
\begin{align*}
\begin{split}
&\bra{\Psi_{1}}H_{\rm coupling}  \ket{ \Psi_{2}} =  i C^2 \frac{\beta (e^{i \phi/2} + e^{-i \phi/2})}{4} \times \int_0^{\infty} dx [ \\
& -4 e^{-\frac{V_z}{ \alpha} x}\cos{(\frac{2m\alpha}{\hbar^2} x)} + 2 e^{-\frac{2\Delta}{ \alpha} x}\cos{(\frac{4 m\alpha}{\hbar^2} x)} + 2 e^{2\frac{\Delta - V_z}{ \alpha} x}]  \\
& = i \frac{ - \beta (e^{i \phi/2} + e^{-i \phi/2})}{2} \frac{\Delta}{V_z} \frac{1 + \frac{5\Delta V_z - 3 V_z^2}{4 \frac{m^2\alpha^4}{\hbar^4}} + \frac{V_z \Delta (2 \Delta - V_z)^2}{16\frac{m^4\alpha^8}{\hbar^8}}}{1 +  \frac{ \Delta^2 + (2\Delta - V_z)^2}{4 \frac{m^2\alpha^4}{\hbar^4}} + \frac{\Delta^2 (2 \Delta - V_z)^2}{16 \frac{m^4\alpha^8}{\hbar^8}}} \\
& \approx i  \frac{\beta \Delta}{V_z}\frac{ e^{i \phi/2} + e^{-i \phi/2}}{2},
\end{split}
\end{align*}
and in the last line we have used the fact that SOC energy is the largest energy scale of the problem.
This produces a Hamiltonian
\begin{equation*} 
H_{\beta} \approx
 \frac{\Delta \beta}{V_z} \left( {
\begin{array}{cc}
0  &  i \frac{e^{i \phi/2} + e^{-i \phi/2}}{2} \\
-i \ \frac{e^{-i \phi/2} + e^{-i \phi/2}}{2} & 0
\end{array} } \right  ),
\end{equation*}
whose eigenenergies are $\pm \frac{\Delta}{V_z} \beta \cos{(\frac{\phi}{2})}$ and they accurately match numerical results presented in Fig.~\ref{fig:analytics_numerics}c.

\vspace{5mm}

\section{Constraints to the scattering matrix due to particle-hole symmetry} \label{app:smat}

We have characterized our system using the determinant of the scattering matrix as the topological invariant. As the scattering matrix is unitary, the absolute value of its determinant has to be one. However, as a result of the constraint the charge-conjugation symmetry imposes on this matrix, this determinant is real and takes two opposite values across the phase transition.

The charge-conjugation symmetry is an anti-unitary symmetry that anti-commutes with the Hamiltonian, such that 
\begin{equation*}
\mathcal{P} H \psi = - \epsilon \mathcal{P} \psi.
\end{equation*} 
Thus, this symmetry can be formally written as $\mathcal{P} = \mathcal{U} \mathcal{K}$, where $\mathcal{U}$ is a unitary operator and $\mathcal{K}$ denotes the complex-conjugation. 
Furthermore, this symmetry also acts on the incoming/outgoing modes present in the scattering process. The incoming modes $\Psi_n^{\rm in}$ are all plane waves with velocity vector pointed into the scattering region, while this vector points outwards for the outgoing modes $\Psi_n^{\rm out}$.  
As the particle-hole symmetry flips energy as well as the momentum vector, the velocity defined as  $\mathbf{v} = \frac{1}{\hbar} \frac{\partial \epsilon}{\partial \mathbf{k}}$ is unaffected and therefore 
\begin{align}
\begin{split} 
& \mathcal{P} \Psi_n^{\rm in} = V^{\mathcal{P}}_{nm} \Psi_m^{\rm in} \\
& \mathcal{P} \Psi_n^{\rm out} =Q^{\mathcal{P}}_{nm} \Psi_m^{\rm out},
\end{split}
\end{align}
where $V^{\mathcal{P}}$ and $Q^{\mathcal{P}}$ are unitary matrices and the summation over the repeated indices in assumed here and in the following.

\begin{figure*}[tb]
 \includegraphics[width=\textwidth]{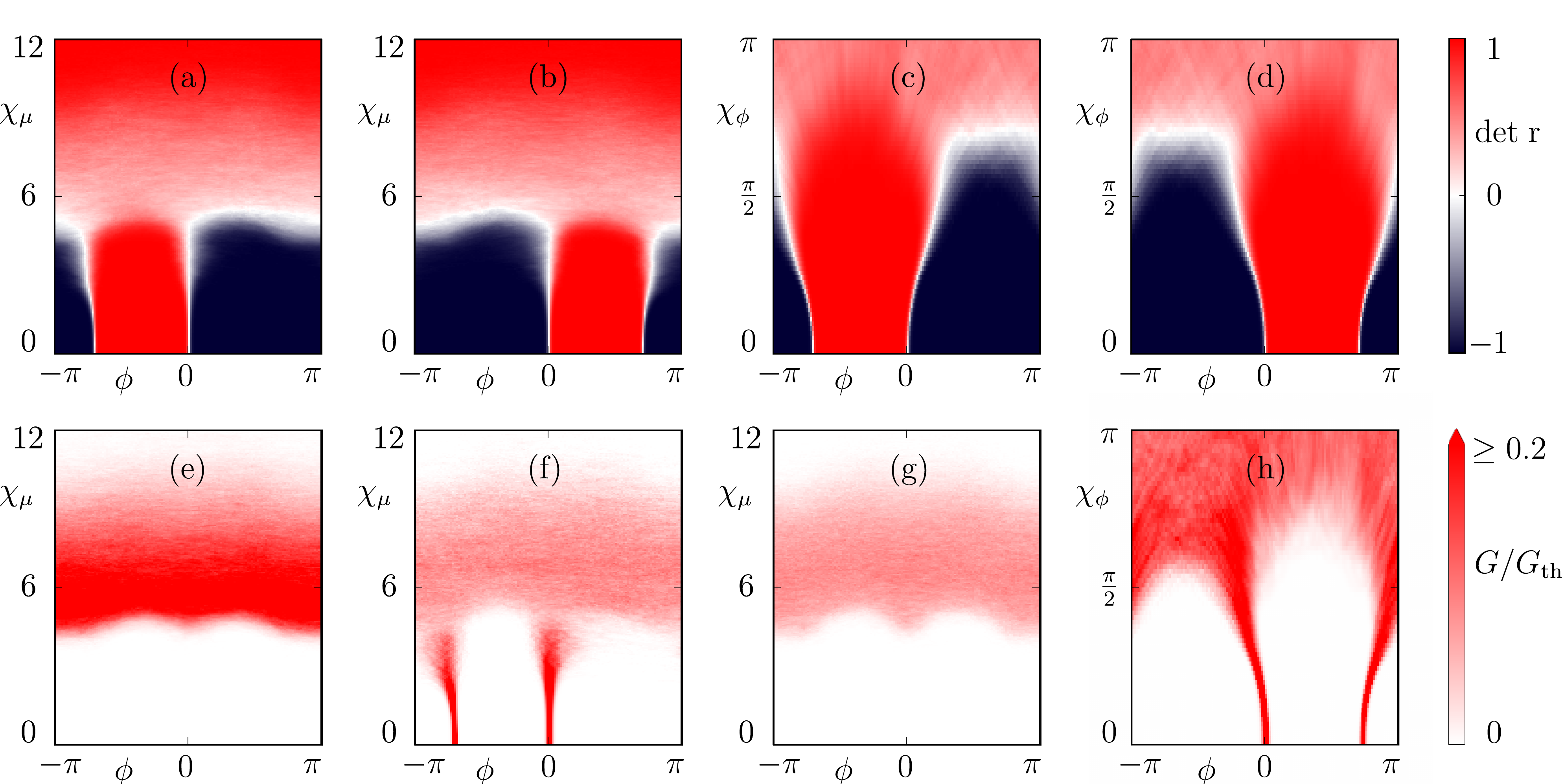}
  \caption{(a-d) Phase diagrams, once there is disorder in the chemical potential $\mu$ (a,b) or in the phases $\phi$ (c,d). Panels (a,c) are obtained from reflection matrices of the left leads, while (b,d) are recovered from reflection matrices of the right leads. These phase diagrams are obtained by averaging topological invariants obtained from 301 disorder realizations.   
(e,f) The averaged value of x-/y-edge conductance, respectively, for disorder in the chemical potential $\mu$ and calculated from the left leads. (g) The bulk conductance calculated using periodic boundary conditions in the y-direction. (h) The averaged y-edge conductance, corresponding to panel (d). In all calculations, the system size is $80 \times 80$ sites for the parameters defined in the main text. }
\label{fig:disorder_Vz}
\end{figure*}

The scattering matrix can be obtained by solving the Schr{\"o}dinger equation
\begin{equation}\label{eq:schrodinger_eq}
(H - \epsilon) (\Psi_n^{\rm in} +  S_{mn} \Psi_m^{\rm out} + \Psi^{\rm loc}) = 0,
\end{equation}
where $\psi^{\rm loc}$ is a wave-function localized near the scattering region. We further simplify the proof by assuming $\Psi^{\rm loc} = 0$, and for the particle-hole symmetric system, $\epsilon = 0$. Then, by applying the charge-conjugation operator on the Eq.~\eqref{eq:schrodinger_eq}:
\begin{align}\label{eq:part_hole_action}
\begin{split} 
& \mathcal{P} H (\Psi_n^{\rm in} +  S_{mn} \Psi_m^{\rm out}) = 0 \rightarrow \\
& - H \mathcal{P}  (\Psi_n^{\rm in} +  S_{mn} \Psi_m^{\rm out}) = 0 \rightarrow \\
& - H  (V^{\mathcal{P}}_{nm} \Psi_m^{\rm in} +  (S^*)_{mn} Q^{\mathcal{P}}_{mp} \Psi_p^{\rm out}) = 0.
\end{split}
\end{align}
Additionally, if $\Psi_n$ satisfies $H \Psi_n = 0$ for any $n$, then the linear
combination $H A_{nm} \Psi_m = 0$. By taking $ A = V^{-1}$ and 
inserting it into the last line of Eq.~\eqref{eq:part_hole_action}
\begin{align} \label{eq:part_hole_schrodinger_eq}
\begin{split}
& H  [ (V^{\mathcal{P}})^{-1}_{q n} V^{\mathcal{P}}_{nm} \Psi_m^{\rm in} +  (V^{\mathcal{P}})^{-1}_{q n} (S^*)_{mn} Q^{\mathcal{P}}_{mp} \Psi_p^{\rm out} ] = 0 \\
&  H  [ \delta_{q m} \Psi_m^{\rm in} +  (V^{\mathcal{P}})^{-1}_{q n} S^{\dagger}_{n m} Q^{\mathcal{P}}_{mp} \Psi_p^{\rm out} ] = 0.
\end{split}
\end{align}
Finally, by comparing Eqs.~\eqref{eq:schrodinger_eq} and \ref{eq:part_hole_schrodinger_eq}, the transformation law for the scattering matrix can be deduced as
\begin{equation*}
S_{pq} = (S^T)_{qp} =  (V^{\mathcal{P}})^{-1}_{q n} S^{\dagger}_{n m} Q^{\mathcal{P}}_{mp}.
\end{equation*} 
By transposing the last equation, one obtains
\begin{equation*}
S =  (Q^{\mathcal{P}})^T S^{*} (V^{\mathcal{P}})^{*}.
\end{equation*} 
Since, in our model $V^{\mathcal{P}} = Q^{\mathcal{P}} = \tau_y \sigma_y$, the scattering matrix obeys 
$ S =  \tau_y \sigma_y S^{*} \tau_y \sigma_y$.

\section{Effects of disorder} \label{app:disorder}
In this section, we study the effects of disorder in the chemical potential and phase difference on the system with the Zeeman field in the z-direction. As this field orientation produces a phase with only two gapless corner modes, we expect that the resulting phase diagrams calculated from right/left leads differ. Furthermore, the simulations of disorder effects on the system with only $V_x$ showed that different system configurations have opposite phase diagrams. Thus, here we study only the first setup (Fig.~\ref{fig:2DSystem}b), as we expect that the phase diagrams obtained from right/left leads of this setup are similar to the phase diagrams obtained from left/right leads of the other setup.
 
As before, fluctuations in the chemical potential are simulated with a variable $v_{\mu}$ that is uniformly distributed in the range $[-\chi_{\mu},\chi_{\mu}]$ and drawn independently for each lattice site. The phase diagrams calculated from left/right leads are presented in Figs. \ref{fig:disorder_Vz}a and  \ref{fig:disorder_Vz}b, respectively.
The x-edge conductance is shown in Fig.~\ref{fig:disorder_Vz}e, while the conductance along left vertical edge is given in Fig.~\ref{fig:disorder_Vz}f. Finally, the bulk conductance is plotted in Fig.~\ref{fig:disorder_Vz}g. Overall, the system behaves similarly to the case of $V_x$ only.

Next, we consider the effects of disorder in the phases of superconductors. This is simulated with a variable  $v_{\phi}$ that is uniformly distributed in the range $[-\chi_{\phi},\chi_{\phi}]$ such that $\phi_{\rm{dis}}(y) = \phi(y) + v_{\phi}$. The resulting phase diagrams are presented in Figs. \ref{fig:disorder_Vz}c and \ref{fig:disorder_Vz}d. The only non-vanishing conductance (on the y-edge), calculated from the left leads, is presented in Fig.~\ref{fig:disorder_Vz}h. As for the other field orientation, disorder in $\phi$ can act detrimentally to the topological phase for one setup while the opposite happens for the other setup. Like before, in the region of small disorder strengths, the topological phase does not expand/reduce significantly.

\bibliography{HOTSC}

\end{document}